\documentclass{pas}
\usepackage{aas_macros}

\usepackage{graphicx}	% Including figure files
\usepackage{adjustbox}  % For the orcid ids
\usepackage{amsmath}	% Advanced maths commands
\usepackage{wasysym}    % Extra astronomy symbols
\usepackage{bm}         % Bold maths
\usepackage{xspace}
\usepackage{xcolor}

\newcommand\HI{\textsc{H\,i}\xspace} % neutral hydrogen
\newcommand\HII{\textsc{H\,ii}\xspace}% ionized hydrogen
\newcommand\K{\text{K}}

\newcommand\dd{\text{d}}

\begin{document}

\lefttitle{Publications of the Astronomical Society of Australia}
\righttitle{Lorinc et al.}

\jnlPage{1}{4}
\jnlDoiYr{2025}
\doival{10.1017/pasa.xxxx.xx}

\articletitt{Research Paper}

\title{Core--wing transitions and the breakdown of diffusion in Lyman-$\alpha$ radiative transfer}

\author{Kevin Lorinc$^{1}$, Aaron Smith$^{1}$, Olof Nebrin$^{2}$, and Joshua Kasiri$^{1}$}

\affil{$^{1}$Department of Physics, The University of Texas at Dallas, Richardson, Texas 75080, USA}

\affil{$^{2}$Department of Astronomy \& Oskar Klein Centre for Cosmoparticle Physics, AlbaNova, Stockholm University, SE-106 91 Stockholm, Sweden }

\corresp{Lorinc. K. Email: \href{mailto:kevin.lorinc@utdallas.edu}{kevin.lorinc@utdallas.edu}}

\begin{abstract}
  The Lyman $\alpha$ (Ly$\alpha$) line of neutral hydrogen plays a central role in observations of star-forming galaxies. However, resonant scattering makes it difficult to directly interpret Ly$\alpha$ signatures. Monte Carlo radiative transfer (MCRT) calculations have become the gold standard for modeling Ly$\alpha$, but it becomes extremely computationally expensive in optically thick environments. Workarounds, such as core-skipping to avoid repetitive low-transport scatterings, greatly increase the efficiency of MCRT simulations but introduce errors in the solutions. While core-skipping is designed to preserve emergent spectra, the internal radiation field, most importantly, the momentum imparted, is not properly preserved. On the other hand, to make analytical and numerical progress, it is often assumed that photons diffuse in both frequency and physical space. We find that these diffusion approximations break down for frequencies near the core and positions at finite optical depths. We propose a more physically-motivated definition for the core--wing transition frequency to isolate such effects. We derive new spectral distributions of internal radiation properties and compare the results with simulations. We analyze the diffusive properties of Ly$\alpha$ photons and demonstrate anomalous spatial diffusion behavior with fat-tailed distributions. This work deepens our understanding of diffusion in resonant-line transfer and identifies areas where simulations or analytics may be failing and how these failures may be resolved.
\end{abstract}

\begin{keywords}
Line: profiles -- radiative transfer -- methods: analytical -- methods: numerical
\end{keywords}

\maketitle

%%%%%%%%%%%%%%%%%%%%%%%%%%%%%%%%%%%%%%%%%%%%%%%%%%

%%%%%%%%%%%%%%%%% BODY OF PAPER %%%%%%%%%%%%%%%%%%

\section{Introduction} \label{sec:intro}

The Lyman-$\alpha$ (Ly$\alpha$) line, corresponding to the $2p \rightarrow 1s$ transition of neutral hydrogen, plays a crucial role in interpreting observations and constraining galaxy formation physics in the high-redshift Universe \citep{Partridge1967}. As hydrogen becomes ionized in \HII regions around young stars, recombination leads to efficient emission of Ly$\alpha$ radiation. Subsequently, Ly$\alpha$ photons undergo a resonant scattering process in the neutral hydrogen surrounding compact ionized bubbles where absorption into the $2p$ state is almost immediately followed by re-emission back into the $1s$ state. This resonant scattering process makes Ly$\alpha$ radiative transfer conceptually and numerically complex, especially in realistic environments with extremely high optical depth \citep{Dijkstra2014}. When benchmarking Ly$\alpha$ radiation pressure calculations, it becomes apparent that frequencies around line center are particularly problematic, highlighting the need to understand its behavior in greater detail. The purpose of this paper is therefore to analyze and decompose the frequency-dependent properties of Ly$\alpha$ radiative transfer to better assess the accuracy and limitations of analytical methods adopting diffusion approximations compared to numerical numerical simulations.

Foundational works motivated the study of Ly$\alpha$ radiative transfer from back-of-envelope scaling arguments, providing physical insights into resonant scattering with significant differences from the properties of monochromatic transfer \citep{Osterbrock1962, Adams1972, Adams1975, HansenOh2006}. Analytical methods have been developed over time, which are mainly based on a diffusion approximation in the full radiative transfer equation for partial frequency redistribution \citep{Unno1952, Hummer1962} rigorously understood within a Fokker--Planck framework \citep{Rybicki1994, Meiksin2006}. The first accurate solutions were for the idealized cases of uniform slab, spherical, and cube geometries \citep{Harrington1973, Neufeld1990, Dijkstra2006, Tasitsiomi2006b}, as well as extended Ly$\alpha$ halos in an expanding neutral intergalactic medium \citet{LoebRybicki1999}. Further generalizations have recently been made, such as analyzing the Wouthuysen-Field effect \citep{Higgins2012, Seon2020}, improvements for boundary conditions and time dependence \citep{McClellan2022}, Ly$\alpha$ feedback during direct collapse black hole formation \citep{GeWise2017, SmithDCBH2017}, methods for Ly$\alpha$ radiation hydrodynamics \citep[][Nebrin et al. in prep.]{Smith2017, Smith2018, Kimm2018, Mushano2024}, and the studies motivating this paper, exploring emissivity, density, destruction, and velocity gradients \citep{LaoSmith2020, Nebrin2025, Smith2025}. 

With the aid of high-performance computing, direct numerical solutions to the radiative transfer equation typically employ Monte Carlo radiative transfer (MCRT), which can be compared with these analytical results for validation before generalizing to more complex geometries \citep{Zheng2002, Dijkstra2006, Tasitsiomi2006, Verhamme2006, Semelin2007, Laursen2009, Yajima2012, Smith2015, GronkeDijkstra2016, Michel-Dansac2020, Seon2020, Byrohl2025}. Most MCRT methods rely on core-skipping algorithms to significantly reduce the computational load of repeated core scatterings, which preserves the emergent spectra but can introduce errors for the internal radiation field. Recently, the dynamical impact of Ly$\alpha$, arising from the momentum transfer that occurs at each scattering event, has been explored as a potentially dominant form of early (i.e. pre-supernova) stellar feedback, especially relevant in the dust-poor conditions of early galaxies \citep{DijkstraLoeb2008, Smith2016, Smith2017, Abe2018, Kimm2018, Tomaselli2021, Mushano2024, Nebrin2025, Ferrara2025, Manzoni2025}. To this end, accurate force calculations are crucial in determining the true dynamical impact of Ly$\alpha$ radiation pressure in such environments.

MCRT provides the gold standard for numerical Ly$\alpha$ radiation transport, since it solves the full radiative transfer equation without any approximations while remaining straightforward to understand and generalize. The MCRT framework makes it possible to track the internal radiation field (e.g. number of scatterings, energy density, momentum exchange, and pressure) and externally observed quantities (e.g. escape fractions, spectral flux, and surface brightness). For internal properties, various biasing techniques can be employed to increase the speed of statistical convergence. The most important ones include differentiating between an event-based sampling approach where photons are treated individually and path-based approaches where photons packets have an associated weight that is updated based on path lengths traversed and the probability of events occurring along the trajectory \citep{Lucy1999, Smith2020}. For properties that are sensitive to the details of scattering, such as the force, an event-based approach can significantly outperform a path-based one, but other properties, such as energy density and dust absorption, greatly benefit from the implicit treatment of changes during transport, so a path-based approach is better. However, even with statistical biasing techniques, MCRT can exhibit poor convergence properties, especially as the optical depth increases, which leads to the need to sample a large number of photon paths before solutions can be fully trusted \citep{Camps2018}. The convergence properties of resonant-line MCRT is further complicated by the frequency-dependent opacity (Kasiri et al., in prep.).

The extreme line-center optical depths make resonant-line Ly$\alpha$ MCRT a computationally expensive method, which led to the development of various acceleration techniques. One of the most common and highly beneficial schemes involves core-skipping algorithms, which preferentially choose the velocity of the scattering atom to push core photons, which are trapped due to extreme optical depths ($\tau_0 \sim 10^6$--$10^{10}$), out into the wing where they can continue their frequency diffusion process \citep{Ahn2002, Smith2015, Michel-Dansac2020}. The speedups can be as large as $\sim 10^3$ times faster, which makes the use of such methods appealing. Alternative speedups can be achieved by resonant Discrete Diffusion Monte Carlo (rDDMC) methods by sacrificing the tracking of exact photon trajectories \citep{Smith2018}. Although external field properties are not significantly affected by the use of acceleration techniques, core skipping can introduce errors in the internal radiation field, especially properties that are sensitive to the total number of scatterings. This means that using any amount of core skipping leads to errors in force calculations, even when the MCRT simulation statistics are fully converged.

As the dynamical importance of Ly$\alpha$ feedback receives additional attention, it is necessary to devise methods where force calculations remain accurate. Fortunately, recent analytical progress in Ly$\alpha$ radiative transfer allows for the stratified development of such methods \citep{LaoSmith2020, Tomaselli2021, McClellan2022, Nebrin2025, Smith2025}. In this paper, we extend these results by deriving the spectral distributions of the force multiplier $M_\text{F}$ and the photon trapping time $t_\text{trap}$ (related to the energy density) for idealized clouds containing neutral hydrogen. We then compare this with the results of MCRT simulations that do not adopt core skipping. This is a robust test of the applicability of the diffusion approximation across the entire spatial--spectral domain of resonant-line transfer. We further explore numerous related numerical experiments to probe how Ly$\alpha$ photons scatter and escape optically-thick environments.

This paper is organized as follows. In Section~\ref{sec:lyman-aplha_rt}, we review the fundamentals of Ly$\alpha$ radiative transfer relevant for this paper, including the MCRT algorithm. In Section~\ref{sec:spectral_distributions} we derive spectral distributions for the force multiplier, the trapping time, and the number of scatterings. We gain insight into the various diffusion approximations, specifically the breakdown of the Fokker--Planck approximation in the core. We introduce a modification to the MCRT algorithms based on the gradient of the energy density to increase the convergence rate of force calculations when core skipping is not used and note that the analytical solution can be used as minor, yet incomplete, correction to the force multiplier when core skipping is used. Finally, we find that the Eddington approximation does not hold everywhere within the cloud, especially near the core--wing transition. In Section~\ref{sec:radiation_properties}, we provide insight into resonant line radiative transfer from robust high-optical depth MCRT simulations with no core skipping, such as the average radius reached and path length traversed before reaching a given frequency for the first time, and we also show the fraction that escapes a certain frequency and compare with the integrated emergent spectra. In Section~\ref{sec:anomalous_diffusion}, we show properties of RT that show anomalous diffusion in fat-tailed distributions along with some statistics about excursions that lead to escape and the probability that a photon at a given frequency returns to the core. In Section~\ref{sec:summary}, we present a summary of our findings and discuss the consequences for numerical algorithms.

\section{Ly$\alpha$ radiative transfer} \label{sec:lyman-aplha_rt}
In this section, we provide some background on Ly$\alpha$ radiative transfer, focusing on aspects relevant for Monte Carlo Radiative Transfer (MCRT) simulations and the theoretical aspects necessary to calculate properties internal to the cloud. During transport, photons probabilistically interact with neutral hydrogen intervening in its path, with a frequency-dependent mean-free path of $\lambda_\text{mfp}(\nu) = 1/n_\HI \sigma_\nu$, where $n_\HI$ is the number density of neutral hydrogen (which we assume is uniform in this paper) and $\sigma_\nu$ is the frequency-dependent cross section. As photons undergo multiple scatterings, they randomly walk in both space and frequency, where the path lengths between scatterings depend on the frequency. To fully describe the steady-state radiation field we need to know the specific intensity $I_\nu(\bm{r},\bm{n})$, which is the energy per unit frequency, $\nu$, and the direction of propagation $\bm{n}$, as a function of spatial position $\bm{r}$. For a static medium with no dust, the steady-state radiative transfer equation is
\begin{equation} \label{eq:general_rt_equation}
  \bm{n} \bm{\cdot} \bm{\nabla} I_\nu = j_\nu - k_\nu I_\nu + \iint k_{\nu'} I_{\nu'} R_{\nu', \bm{n}' \rightarrow \nu, \bm{n}} \, \text{d}\Omega' \text{d}\nu' \, ,
\end{equation}
where $j_\nu$ and $k_\nu$ are the emission and absorption coefficients, respectively. The integral in the last term on the right accounts for the frequency redistribution caused by partially coherent scattering \citep[e.g.][]{Dijkstra2014}. The conditional redistribution function $R_{\nu', \bm{n}' \rightarrow \nu, \bm{n}}$ is the differential probability per unit initial photon frequency $\nu'$ and per unit initial directional solid angle $\Omega'$ that scattering of such a photon traveling in the direction $\bm{n}'$ would place the scattered photon at frequency $\nu$ and in the direction of the unit vector $\bm{n}$. The development of resonant scattering redistribution functions is given in detail in \citet{Hummer1962} and \citet{Hubeny2015}. For Ly$\alpha$ the appropriate redistribution function is $R_\text{II}$, but notice the way we define it here that it is a conditional probability rather than a joint probability. This makes for easier analytical progress. In treating single-line problems like Ly$\alpha$, it is convenient to switch to the dimensionless frequency
\begin{equation} \label{eq:x}
  x \equiv \frac{\nu - \nu_0}{\Delta \nu_\text{D}} \, ,
\end{equation}
where $\nu_0 = 2.466 \times 10^{15} \, \text{Hz}$ is the frequency at line center, $\Delta \nu_\text{D} \equiv (v_\text{th}/c)\nu_0$ the Doppler width of the profile, and $v_\text{th} \equiv (2 k_\text{B} T / m_\text{H})^{1/2}$ the thermal velocity. To characterize the frequency dependence on the cross section, we define the Hjerting--Voigt function $H(a,x) = \sqrt{\pi} \Delta \nu_\text{D} \phi_\text{Voigt}(a, \nu)$ as the dimensionless convolution of Lorentzian and Maxwellian distributions,
\begin{equation} \label{eq:H}
  H(a,x) = \frac{a}{\pi} \int_{-\infty}^\infty \frac{e^{-y^2}\text{d}y}{a^2+(y-x)^2} \approx
    \begin{cases}
      e^{-x^2} & \quad \text{`core'} \\
      {\displaystyle \frac{a}{\sqrt{\pi} x^2} } & \quad \text{`wing'}
    \end{cases} \, .
\end{equation}
The `damping parameter', $a \equiv \Delta \nu_{\rm L} /2 \Delta \nu_{\rm D}$, describes the relative thermal broadening compared to the natural line width. We focus on the case of Ly$\alpha$ where $\Delta \nu_\text{L} = 9.936 \times 10^7 \, \text{Hz}$, and, in an isothermal gas, the damping parameter traces the temperature $T$. Specifically $a \approx 4.702 \times 10^{-4} \, (T/10^4 \, \K)^{-1/2}$. The cross section at dimensionless frequency $x$ is $\sigma_x = \sigma_0 H(a,x)$, where $\sigma_0 = 5.898 \times 10^{-14} \, (T/10^4 \, \K)^{-1/2} \, \rm cm^2$ is the cross section at line center. This allows us to give the optical depth as
\begin{equation} \label{eq:tau}
    \tau_x = \int_\text{path} k(\bm{r}) \, H(a,x) \, \text{d}\ell \, ,
\end{equation}
where $k(\bm{r}) = n_\HI(\bm{r}) \sigma_0$ is the absorption coefficient at line center at position $\bm{r} = \bm{r}_0 + \ell \bm{n}$. We define the line center optical depth $\tau_0$ as the integrated value from the center to the outer boundary of a spherical cloud, and use this throughout as a simulation parameter.

In general, it is only possible to solve Eq.~(\ref{eq:general_rt_equation}) numerically, however further analytical progress can be made if the product $a \tau_0$ is sufficiently large \citep[for a comprehensive discussion see][]{Nebrin2025}. In this regime, scattering may be treated as a diffusion process in both space and frequency, and we find analytical solutions for the angular-averaged frequency $J_x \equiv \frac{1}{4\pi} \int I_x \, \text{d} \Omega$. To do so, we integrate Eq.~(\ref{eq:general_rt_equation}) further defining $\bm{H}_x \equiv \frac{1}{4\pi} \int \bm{n} I_x \, \text{d} \Omega$ to give us
\begin{align} \label{eq:angular_avg_rt_equation}
    \ \bm{\nabla} \bm{\cdot}  \bm{H}_x = \int \frac{j_x}{4\pi} \, \text{d} \Omega - k_x J_x + \int k_{x'} J_{x'} R_{x' \rightarrow x} \, \text{d}x' \, ,
\end{align}
where $R_{x' \rightarrow x} \equiv (4\pi)^{-2} \iint \text{d}\Omega' \text{d}\Omega \, R_{x', \bm{n}' \rightarrow x, \bm{n}}$ is the angular-averaged redistribution function, with $R_{x' \rightarrow x} = \Delta\nu_\text{D}^2 \, R_{\nu' \rightarrow \nu}$, and $J_x = \Delta\nu_\text{D} J_\nu$. In optically thick environments, the intensity is nearly isotropic (Eddington approximation), and we can apply Fick's law as a closure relation to the moment equations:
\begin{equation} \label{eq:Ficks_Law}
  \bm{H}_x \approx -\frac{\bm{\nabla} J_x}{3 k_x} \, .
\end{equation}
The Fokker--Planck approximation (FPA) is often used to approximate the redistribution integral, but, as will be explained below, the common form of the approximation is only accurate for wing frequencies \citep{Rybicki1994,Rybicki2006, McClellan2022} where frequency diffusion occurs. The error of the FPA is explicitly shown in the Appendix~\ref{sec:rf_moments} from which it is clear that the Fokker--Planck treatment can only be expected to have good results in the wing. The usual approximation is
\begin{align} \label{eq:fokker-planck}
    -k_xJ_x + &\int k_{x'} J_{x'} R_{x' \rightarrow x} \, \text{d}x' \approx
    \frac{1}{2}\frac{\partial}{\partial x}\left(k_x\frac{\partial J_x}{\partial x}\right) \, ,
\end{align}
from which one can write down the diffusion equation under an analytical transform and solve for $J$. Then one can derive the radiation energy density, related to the angular-averaged intensity as $u(\bm{r}) = \frac{4\pi}{c} \int J_x(\bm{r})\,\text{d}x$, the trapping time, defined as $t_\text{trap} \equiv \mathcal{L}^{-1} \int u(\bm{r})\,\text{d}V$ with $\mathcal{L}$ the source luminosity,\footnote{The trapping time is often normalized by the light crossing time, which has geometric correction factors for extended sources \citep{Smith2025}.} and the outward force multiplier
\begin{equation} \label{eq:M_F_def}
  M_\text{F}
  \equiv \mathcal{L}^{-1}  \iint k(\bm{r}) F \,\text{d}x\,\text{d}V 
  \approx -\frac{c}{3\mathcal{L}}  \int \nabla u(\bm{r})\,\text{d}V \, ,
\end{equation}
where the second expression follows from Ficks law, replacing the flux by $F \approx -c \nabla u / 3k$ for diffusive radiation. The force multiplier quantifies the enhancement of the momentum coupling compared to the single scattering limit of $\mathcal{L}/c$ \citep[e.g.][]{DijkstraLoeb2008}. We define the Ly$\alpha$ scattering rate as $P_\alpha = (4\pi / \mathcal{L})\int \sigma_x J_x \, \dd x$, where we normalize by the total luminosity $\mathcal{L}$ since we express intensity in terms of energy instead of photon number, as previous authors have done in \citet{Higgins2012} and \citet{Seon2020}. Then the average number of scatterings is $N_\text{scat} = \int P_\alpha n_\HI\,\dd V$ which becomes
\begin{equation} \label{eq:n_scat_def}
     N_\text{scat} = \frac{4 \pi}{\mathcal{L}} \iint k(\bm{r})\,H(a,x)\,J_x(\bm{r})\,\dd x\,\dd V \, .
\end{equation}

The diffusion approximation has been analytically explored in numerous works \citep[e.g.][]{Harrington1973, Neufeld1990, Dijkstra2006, Tomaselli2021, McClellan2022}, however, the notation and definitions in this paper are consistent with \citet{LaoSmith2020} and \citet{Smith2025}, which explore more general solutions that account for gradient structures in density and velocity space, respectively. In \citet{Nebrin2025}, a combination of a wide range of effects were analytically treated. All analytical results were validated with MCRT simulations, and the diffusion approximation is accurate as long as $a\tau_0 \gg 1000$.

To directly solve Eq.~(\ref{eq:general_rt_equation}) without any approximations, we employ the MCRT algorithm which samples a large number of individual photon paths to generate statistically converged properties of the radiation field. We utilize the Cosmic Ly$\alpha$ Radiative Transfer code (\textsc{colt}), which is described in \citet{Smith2015, Smith2019, SmithMW2022} and \citet{McClymont2025}. Concisely, the optical depth to be traversed between scattering events is drawn from an exponential distribution, consistent with the interpretation of the optical depth $\tau_x$ from Eq.~(\ref{eq:tau}) as the inverse mean free path between photon interactions. Once the photon has traversed enough distance to attenuate this randomly drawn optical depth, the photon undergoes a scattering event, and the frequency and direction are updated accordingly. This process is repeated until the photon escapes the computational domain. Sampling enough photon paths will result in converged properties of the radiation field thus solving Eq.~(\ref{eq:general_rt_equation}) for the intensity $I(\bm{r},x,\bm{n})$ and related quantities with the definitions provided.

Although accurate when fully converged without using any acceleration scheme, resonant-line MCRT suffers significantly from repeated core scatterings when the photon is near the line center frequency, which causes the algorithm to slow down. Each time the photon goes to a resonant frequency, it gets trapped in space until a random excursion into the wing of the line (caused by an interaction with a hydrogen atom of sufficiently large velocity) enables the photon to move a non-negligible distance. Further, the error scales linearly with optical depth, meaning that higher-density simulations would require more photons for the same level of error, contributing to a very long runtime to produce converged simulations. The exact details of Ly$\alpha$ MCRT convergence are given in (Kasiri et al., in prep). To avoid photon trapping, the core-skipping algorithm picks an interaction with an atom that has enough velocity to push the core photons out into the wing of the line, the boundary of which will be defined in Section~\ref{sec:spectral_distributions}. Core-skipping retains accuracy for emergent properties, since photons cannot move in physical space until they diffuse in frequency and are already in the wing. Unfortunately, core photons can considerably impact internal properties, specifically the force multiplier, since repeated core scatterings impart momentum in a way that does not cancel out. Clearly, the inherent accuracy of Monte Carlo is challenged by core-skipping, especially for frequency-dependent quantities, which we explore in this paper as the cause for missing momentum when comparing analytical results to Monte Carlo results. To avoid misinterpreting MCRT simulations, we refrain from using core-skipping in our simulations unless otherwise mentioned.

\section{Spectral distributions of radiation}
\label{sec:spectral_distributions}
\begin{figure}
    \centering
    \includegraphics[width=\columnwidth]{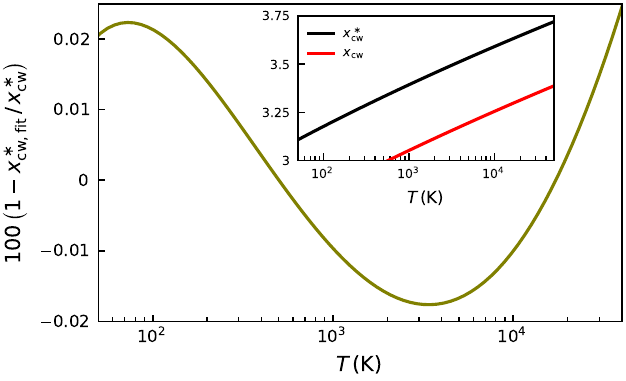}
    \caption{The relative percent error in the fitting formula given by Eq.~(\ref{eq:xcwast_fit}) is shown in green. The bound on the error is very low for the parameter range of interest. In the inset plot, the exact solution for largest solution to Eq.~(\ref{eq:force_core_wing_frequency}) found with a root finder is shown in black while the red line indicates $x_{\text{cw}}$ defined as the frequency where $H(a,x)$ transitions from a Gaussian to a Lorentzian distribution, which is important for core-skipping algorithms.}
    \label{fig:xcwast}
\end{figure}
Core photons can contribute significantly to internal radiation properties, such as the total number of scatterings, the trapping time, and the total momentum imparted. Fully capturing these contributions in simulations requires running MCRT simulations without the commonly adapted core-skipping algorithms that result in a significant ray-tracing speed-up. From previous analytical developments in Ly$\alpha$ RT \citep{LaoSmith2020, Nebrin2025, Smith2025}, we can derive spectral distributions of internal fields to quantify the relative contribution of core scatterings relative to wing scatterings based on the diffusion approximation for the various internal quantities. We use the results from \citet{LaoSmith2020} in reference to spherical geometry without any velocity or density gradients. We compare these distributions with MCRT simulations that do not use core-skipping to analyze the accuracy of our analytical findings. Using core-skipping gives results that converge to the wrong answer because of the missing core contributions. To amend this, a semi-analytical core correction can approximate core contributions to the force multiplier, trapping time, and number of scatterings while allowing for the use of core skipping. This works quite well for the trapping time and the number of scatterings, but the force multiplier suffers from the breakdown of diffusion approximations, which we elaborate on. More practically, we show that the force can be efficiently calculated from a quickly convergent energy density or pressure calculation with even less error than the analytical correction. We compare this with an upcoming Ly$\alpha$ code based on the diffusion approximation, \textsc{lydion-1d} (Nebrin et al., in prep). Future calculations rely heavily on $\tilde{x}$, defined in differential form as
\begin{equation} \label{eq:dx_tidle}
    \text{d}\tilde{x} = \sqrt{\frac{2}{3}}\frac{\text{d}x}{\tau_0 H(a, x)} \, .
\end{equation}
Here $\tilde{x}$ depends on both the temperature and frequency through $H(a,x)$. When finding exact solutions, we use numerical integration to solve for $\tilde{x}(x)$. To distinguish between the ``core'' and ``wing'' frequencies, we use the approximations in Eq.~(\ref{eq:H}) to find
\begin{equation} \label{eq:x_tilde_approx}
    \text{Wing: }\, \tilde{x} \approx \sqrt{\frac{2\pi}{27}} \frac{x^3}{a \tau_0} \quad \text{Core: } \,\tilde{x} \approx \sqrt{\frac{\pi}{6}} \frac{\text{erfi}(x)}{\tau_0} \,,
\end{equation}
where the imaginary error function is $\text{erfi}(x) = -i\,\text{erf}(i x)$. Equating the two approximations and solving gives a critical frequency ${x_\text{cw}^\ast}$. This frequency is different from $x_\text{cw}$ from eq.~(21) of \citet{Smith2015}, which is defined by equating the two approximations in Eq.~(\ref{eq:H}) for $H(a,x)$. We show $x_\text{cw}^\ast$, defined by equating the approximations after integrating $1/H(a,x)$, better characterizes differences in behavior between the core and the wing than the traditional $x_\text{cw}$. The transcendental equation for $x_\text{cw}^\ast$ is
\begin{equation} \label{eq:force_core_wing_frequency}
    \frac{2}{3} {x_\text{cw}^\ast}^3 = a \,\text{erfi}({x_\text{cw}^\ast}) \, ,
\end{equation}
and an accurate fitting formula in our relevant parameter range is
\begin{equation} \label{eq:xcwast_fit}
    {x_\text{cw}^\ast}(a) = \frac{1.53760 - 0.37024 \ln(a)}{0.85538 - 0.04730 \ln(a)} \, .
\end{equation}
We plot the relative error, which is strongly bounded over the relevant parameter range, in Fig.~\ref{fig:xcwast} and show the exact form in an inset versus the fitting formula along with the value of $x_\text{cw}$ as a reference in red to highlight the difference in definition. The procedure through this section is as follows: let $Q$ be a general radiative transfer quantity, e.g. from Eqs.~(\ref{eq:M_F_def})--(\ref{eq:n_scat_def}). We define the spectral distribution of $Q$ as the frequency derivative $\dd Q/\dd x \equiv Q_x$. Similarly, the contribution to $Q$ over a frequency range $x_\text{min}$ to $x_\text{max}$ is $\int_{x_\text{min}}^{x_\text{max}} Q_x \dd x$. To find the core contribution, $Q^\text{core}$, take $x_\text{max} = x_\text{cw}^\ast$ and $x_\text{min} = - x_\text{cw}^\ast$. Then the wing contribution is defined as $Q^\text{wing} \equiv Q - Q^\text{core}$. The fraction of contribution to $Q$ coming from frequencies $<|x|$ is defined as
\begin{equation} \label{eq:fractional_contribution}
    % f_{Q,<|x|} \equiv \int_{-x}^x Q_x \,\dd x \, \bigg/ \int_{-\infty}^{\infty} Q_x \,\dd x \, ,
    f_{Q,<|x|} \equiv \frac{\int_{-x}^x Q_x \,\dd x}{\int_{-\infty}^{\infty} Q_x \,\dd x} \, ,
\end{equation}
and the label can vary depending on $Q$, but is descriptive of the quantity.

\begin{figure}
    \centering
    \includegraphics[width=\columnwidth]{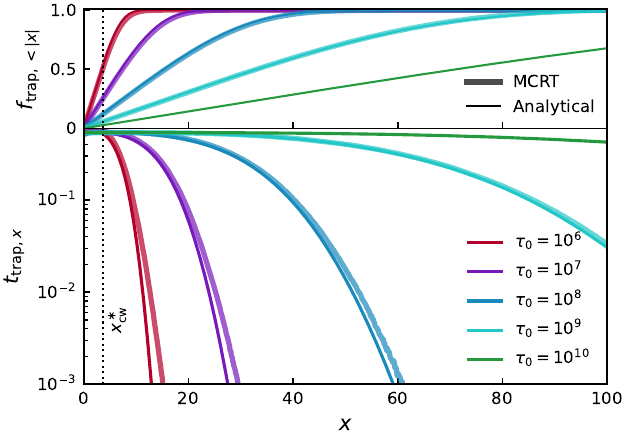}
    \caption{A comparison between analytical trapping time (thin lines) and MCRT simulations (thick lines) as defined in Eq.~(\ref{eq:general_spectral_trapping_time}). \textbf{Bottom Panel:} The analytical $t_{\text{trap},x}$ from Eq.~(\ref{eq:spherical_point_spectral_t_trap}) evaluated exactly with numerical integration for various values of $\tau_0$ with a fixed temperature of $T = 10^4 \,\K$ compared with binned contributions from MCRT simulations. The vertical line shows $x_\text{cw}^\ast$ at this temperature. \textbf{Top Panel:} The cumulative fraction of $t_\text{trap}$ from frequencies $< |x|$ ($f_\text{trap}$) as defined in Eq.~(\ref{eq:fractional_contribution}). The agreement here is almost perfect once the diffusion approximation is valid.}
    \label{fig:t_trap_x}
\end{figure}

\subsection{Spectral trapping time}
The spectral trapping time in a general finite-cloud geometry is defined as
\begin{equation} \label{eq:general_spectral_trapping_time}
    t_{\text{trap},x} = \frac{\dd t_{\text{trap}}}{\dd x} \equiv \frac{4 \pi}{\mathcal{L} c} \int J_x(\bm{r}) \,\dd V \, .
\end{equation}
We define volume-weighted averages as 
\begin{equation} \label{eq:volume_weighted_average}
  \langle f \rangle \equiv \frac{\int f(\bm{r})\,\text{d}V}{\int \text{d}V} \, .
\end{equation}
In spherical geometry, with a total cloud radius $R$ we have $t_{\text{trap},x}/t_\text{light} = (16 \pi^2 R^2 / 3\mathcal{L})\,\langle J_x \rangle$, where $\langle J_x \rangle$ is the average internal spectrum at a given frequency $x$, and the light crossing time is $t_\text{light} \equiv R/c$. For a static uniform density spherical point source cloud this simplifies to (see Eq.~90 in \citealt{LaoSmith2020})
\begin{equation} \label{eq:spherical_point_spectral_t_trap}
    \frac{t_{\text{trap},x}}{t_\text{light}} = \frac{\sqrt6}{\pi} \ln\left(1 + e^{-\pi |\tilde{x}|}\right) \, .
\end{equation}
To find the trapping time contributions from various frequency ranges, we must integrate the spectral trapping time over a frequency range. We consider the two cases by defining the core contribution as the quantity integrated over core frequencies ($|x| \leq x_\text{cw}^\ast$), and the wing contribution as the quantity integrated over wing frequencies ($|x| > x_\text{cw}^\ast$). Under the core approximation,
\begin{align}
    t_{\text{trap},x}^\text{core}\,/\,t_\text{light} &= \frac{\sqrt6}{\pi} \ln\left[1 + \exp\left(-\sqrt{\frac{\pi^3}{6}} \frac{\text{erfi}(|x|)}{\tau_0}\right)\right] \notag \\
    &\approx \frac{\sqrt6}{\pi} \ln(2) \approx 0.540 \, ,
\end{align}
where we make use of the fact that $x < {x_\text{cw}^\ast} \implies \text{erfi}(x)/\tau_0 \ll 1$ for any physical optical depth under the diffusion regime. Thus, after integrating from $-{x_\text{cw}^\ast}$ to ${x_\text{cw}^\ast}$, we find $t_{\text{trap}}^\text{core} \,/\, t_\text{light} = (2\sqrt{6}/\pi \,\ln(2)) \, {x_\text{cw}^\ast}$, which only varies weakly with temperature. Under the wing approximation,
\begin{equation}
    t_{\text{trap},x}^\text{wing}\,/\,t_\text{light} = \frac{\sqrt6}{\pi} \ln\left[1 + \exp\left(-\sqrt{\frac{2\pi^3}{27}} \frac{|x^3|}{a\tau_0}\right)\right] \, ,
\end{equation}
which may be asymptotically expanded to get a far wing approximation but needs to be numerically integrated to find the general wing contribution. In the limit as $x \rightarrow 0$, both the core and wing approximation approach the same value, so the overall spectral distribution of the trapping time is well described by the wing approximation for all frequencies while the core contribution quickly dies off for $|x| > {x_\text{cw}^\ast}$. In the bottom panel of Fig.~\ref{fig:t_trap_x}, we plot the spectral trapping time from Eq.~(\ref{eq:spherical_point_spectral_t_trap}) with thin lines, and MCRT results with thick lines. In the top panel of Fig.~\ref{fig:t_trap_x}, we integrate the spectral distribution giving the contribution from the trapping time coming from frequencies $<|x|$. The agreement is excellent and the error decreases as $\tau_0$ increases. As expected, the distribution in the core is roughly constant at all optical depths, and the wing spread increases as the optical depth increases, which leads to a broader emission line. This follows from increased diffusion as the medium becomes more opaque for wing photons. The wing contribution quickly beats out the core contribution as $\tau_0$ increases.

\begin{figure}
    \centering
    \includegraphics[width=\columnwidth]{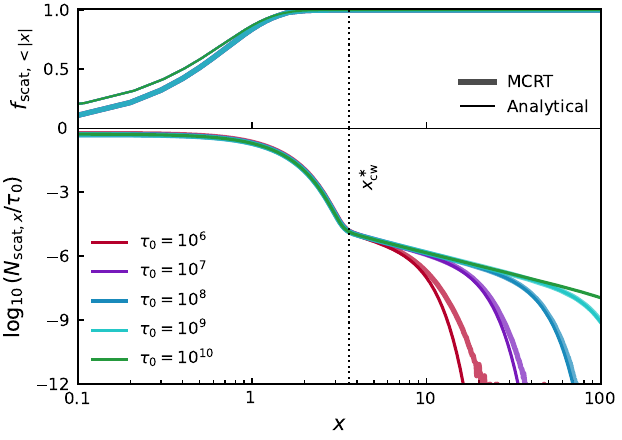}
    \caption{A comparison between the analytical spectral number of scatterings (thin lines) and MCRT simulations (thick lines) as defined in Eq.~(\ref{eq:general_N_scat_x}). \textbf{Bottom Panel:} The analytical $N_{\text{scat},x}$ from Eq.~(\ref{eq:N_scat_x_point}) normalized by $\tau_0$ as a function of $x$ at various $\tau_0$ at a fixed temperature $T = 10^4 \, \K$ compared to binned MCRT simulation data. The vertical dotted line shows the value of $x_\text{cw}^\ast$ for this temperature. \textbf{Top Panel:} The cumulative scattering fraction $f_{\text{scat},<|x|}$ as defined in Eq.~(\ref{eq:fractional_contribution}).}
    \label{fig:n_scat_x}
\end{figure}

\subsection{Spectral number of scatterings}
By a similar notion, we can analyze the spectral average scattering value by defining
\begin{equation} \label{eq:general_N_scat_x}
    N_{\text{scat},x} \equiv \frac{\dd N_\text{scat}}{\dd x} = \frac{4\pi}{\mathcal{L}} \int J_x(\bm{r}) k(\bm{r}) H(a,x) \, \dd V \, ,
\end{equation}
which in the case of a constant density static spherical cloud, using Eq.~(\ref{eq:spherical_point_spectral_t_trap}), can be expressed in terms of the spectral trapping time as
\begin{equation} \label{eq:N_scat_x_point}
    N_{\text{scat},x} = c k_0 H(a,x)\,t_{\text{trap},x} = \frac{\sqrt6}{\pi} \tau_0\,H(a,x) \ln\left(1 + e^{-\pi|\tilde{x}|}\right) .
\end{equation}
In Fig.~\ref{fig:n_scat_x}, the bottom panels plot the spectral distribution of the number of scatterings while the top panel plots the cumulative contribution defined in Eq.~(\ref{eq:fractional_contribution}) for a variety of optical depths against MCRT simulations. The agreement with MCRT simulations is very good, with only a small error near the core. Across different optical depths, the total $N_\text{scat}$ is fully determined by the core scatterings. There is excellent agreement in the core and, as will be predicted by Eq.~(\ref{eq:N_scat_core}), when normalized by $\tau_0$, the distributions lie on top of each other. The cumulative plots are all on top of each other because of the negligible contribution of wing scatterings to the total number of scatterings. The far wing becomes noisy as the number of photons that undergo a lot of scatterings is small, especially for smaller $\tau_0$. To further analyze the core and wing contributions and their scaling, we proceed with analytics. In previous work \citep{Higgins2012, Seon2020}, $H(a,x)$ has been treated as a delta function centered at $x = 0$ with normalization $\int H(a,x) f(x) \dd x \approx \sqrt\pi f(0)$, from which we recover the usual result for the average number of scatterings in a sphere:
\begin{equation} \label{eq:N_scat_delta}
    N_{\text{scat},\delta} \approx \left(\sqrt\frac{6}{\pi} \ln2 \right) \tau_0 \, \approx 0.958 \, \tau_0 \, ,
\end{equation}
where $\delta$ signifies that $H(a,x)$ was treated here as a delta function. This assumes that all scatterings occur exactly at line center, which is true only when $\tau_0 \rightarrow \infty$. For more precision at intermediate optical depths (where the diffusion approximation still holds) and to gain insight into scalings, we derive the core and the wing contributions. Expanding $H(a,x)$ to second order in $a$ gives
\begin{equation}\label{eq:H_expanded}
    H(a, x) \approx e^{-x^2} + \frac{2 a}{\sqrt{\pi}} [ 2 x F(x) - 1] + a^2 e^{-x^2} ( 1 - 2 x^2 ) \, ,
\end{equation}
where $F(x) \equiv \int_0^x e^{y^2 - x^2} \dd y$ is the Dawson function. We define the first frequency integral of the Hjerting--Voigt function
\begin{equation} \label{eq:Upsilon}
    \Upsilon(a,x) \equiv \int H(a,x)\,\text{d}x \approx \frac{\sqrt{\pi}}{2} \text{erf}(x) - \frac{2 a}{\sqrt{\pi}} F(x) + a^2 x e^{-x^2} \, ,
\end{equation}
as in Eq.~(82) of \citet{Smith2025}. Using the core expression for $t_\text{trap}^\text{core}$, we find that for large optical depths
\begin{align} \label{eq:N_scat_core}
    N_\text{scat}^\text{core} &= 2\int_0^{x_\text{cw}^\ast}\frac{\sqrt6}{\pi} \tau_0 H(a,x)  \ln\left[1 + \exp\left(-\sqrt{\frac{\pi^3}{6}} \frac{\text{erfi}(|x|)}{\tau_0}\right)\right]  \dd x\notag \\
    &\approx \left(\frac{2\sqrt{6}}{\pi} \ln(2) \Upsilon(a, x_\text{cw}^\ast)\right) \tau_0  \, .
\end{align}
Treating $H(a,x)$ as an effective delta function is equivalent to assuming that all scatterings occur in the core, so we can recover Eq.~(\ref{eq:N_scat_delta}) from Eq.~(\ref{eq:N_scat_core}) by evaluating $\Upsilon(x_\text{cw}^\ast)$ as $x_\text{cw}^\ast \rightarrow \infty$ (only the zeroth order term survives, and $\text{erf}(x) \rightarrow 1 \,\text{as}\, x \rightarrow \infty$). We note that for most physical temperatures, Eq.~(\ref{eq:N_scat_delta}) is quite accurate to the zeroth order. For the wing limit, we find
\begin{align} \label{eq:N_scat_wing}
    N_\text{scat}^\text{wing} &= 2 \int_{x_\text{cw}^\ast}^\infty \frac{\sqrt6}{\pi} \tau_0 H(a,x) \ln\left[1 + \exp\left(-\sqrt{\frac{2\pi^3}{27}} \frac{|x^3|}{a\tau_0}\right)\right] \dd x \notag \\
    &\approx \frac{2\sqrt6}{\pi^{3/2}} a\tau_0 \int_{x_\text{cw}^\ast}^\infty \exp\left(-\sqrt{\frac{2\pi^3}{27}} \frac{|x^3|}{a\tau_0}\right)\bigg/x^2 \, \dd x \notag \\ 
    &= \frac{2}{\sqrt3 \pi} \left(\frac{2}{27}\right)^{1/6} \Gamma \left(-\frac{1}{3},\sqrt{\frac{2\pi^3}{27}} \frac{{x_\text{cw}^\ast}^3}{a \tau_0} \right) (a\tau_0)^{2/3}\, ,
\end{align}
where $\Gamma(z,x) = \int_x^\infty t^{z-1}e^{-t}\dd t$ is the upper incomplete gamma function. The integrand is negligible compared to $\tau_0$, so the wing contribution will be negligible compared to the number of core scatterings, i.e., $N_\text{scat}^\text{wing} \ll N_\text{scat}^\text{core}$. Generally, $x_\text{cw}^\ast \ll (a \tau_0)^{1/3}$ in the diffusion limit, so we may Taylor expand the incomplete gamma function,
\begin{equation}
    \Gamma \left(-\frac{1}{3},\sqrt{\frac{2\pi^3}{27}} \frac{x_\text{cw}^3}{a \tau_0} \right) \approx \Gamma\left(-\frac{1}{3}\right) + \frac{3}{x_\text{cw}^\ast} \left(\frac{2 a \tau_0}{3}\right)^{1/3} \, .
\end{equation}
Note that $\Gamma(-1/3) \approx -4.06 < 0$. Then the wing scatterings have a term $\propto \tau_0$, indicating the number of returns from the core (via wing excursions with scatterings $\sim$ of order unity) and a negative correction $\propto (a\tau_0)^{2/3}$ corresponding to the final wing excursion leading to escape:
\begin{equation} \label{eq:N_scat_wing_approx}
    N_\text{scat}^\text{wing} \approx \frac{2}{\sqrt3 \pi} \left(\frac{2}{27}\right)^{\frac{1}{6}} \Bigg[\!\underbrace{\Gamma\left(-\frac{1}{3}\right) (a\tau_0)^{\frac{2}{3}}}_\text{final wing excursion} + \underbrace{\frac{3}{x_\text{cw}^\ast} \left(\frac{2}{3}\right)^{\frac{1}{3}} (a\tau_0)}_\text{core-wing transitions}\Bigg] \, .
\end{equation}
Note that $x_\text{cw}^\ast$ depends on $a$ in a nontrivial manner (see Eq.~\ref{eq:xcwast_fit}), canceling out some of the temperature dependence; however, the scaling with $\tau_0$ remains unchanged. More optically thick clouds need more wing scatterings to get excursions far enough into the wing for the opacity to drop off and for the photon to escape. However, the scaling differs between different types of wing scatterings between excursions that drift back to the core and excursions that lead to escape. The $\tau_0$ scaling is due to small excursions into the wing that drift back to the core. These are much more common than lucky excursions that take you far enough from line center for escape. The relative dominance of return to the core over these rare wing excursions is easily understood by the exponential-Lorentzian decay of the redistribution function $R_{\text{II}}$ for core-wing transitions \citep[e.g.][]{Hummer1962, Ayres1985}, which falls off asymptotically as $e^{-x^2/4}/x^2$ when the incoming frequency $x' < x_\text{cw}^\ast$. The $(a\tau_0)^{2/3}$ scaling is the correction for wing scatterings in the final wing excursion leading to escape and follows from back of the envelope estimates of $N_\text{scat} \sim \tau_\text{crit}^2$, with the optical depth at the critical escape frequency being $\tau_\text{crit} \sim x_\text{crit} \sim (a\tau_0)^{1/3}$.

\begin{figure}
    \centering
    \includegraphics[width=\columnwidth]{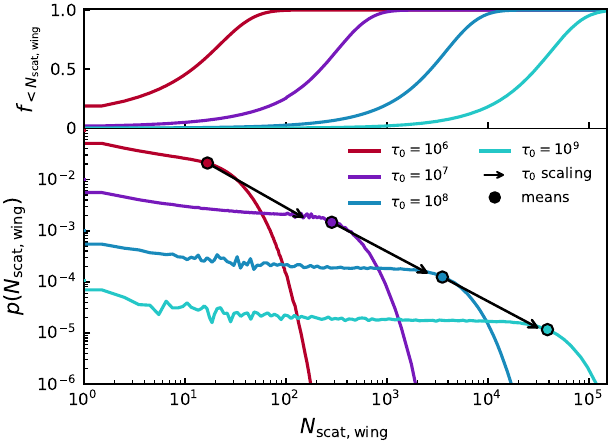}
    \caption{Statistics on the number of wing scatterings for photons that eventually return to the core. \textbf{Bottom Panel:} The probability density function of photons having $N_\text{scat, wing}$ scatters in the wing excluding the scatters in the final excursion leading to escape for various values of $\tau_0$. The temperature is fixed at $T = 10^4 \, \K$. The dots represent the mean of the distribution and the arrows point from one mean to the location of the next predicted mean based off the analytical form in Eq.~(\ref{eq:N_scat_wing_approx}). The scaling comes to better agreement as the optical depth increases. \textbf{Top Panel:} The cumulative probability that a photon scatters less than $N_\text{scat,wing}$ times prior to the final wing excursion.}
    \label{fig:n_scat_wing_dist}
\end{figure}

\begin{figure}
    \centering
    \includegraphics[width=\columnwidth]{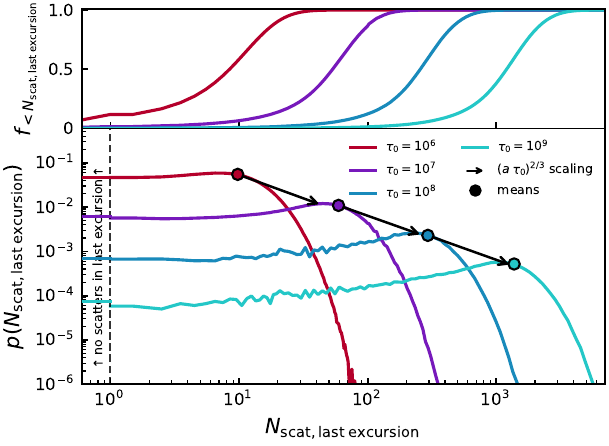}
    \caption{Statistics on the number of wing scatterings for photons that eventually return to the core. \textbf{Bottom Panel:} the probability density function of photons having $N_\text{scat, last excursion}$ scatterings in the final wing excursion leading to escape across several optical depths. The temperature is fixed at $T = 10^4 \,\K$. Again, the dots present means and the arrows point to the next predicted mean based off the scaling in Eq.~(\ref{eq:N_scat_wing_approx}). The agreement is excellent at higher optical depths. The vertical dotted line distinguishes the leftover photons in the zero bin that are plotted discretely. \textbf{Top Panel:} the cumulative distribution giving the probability that a photon has $< N_\text{scat, last excursion}$ scatterings in its last excursion.}
    \label{fig:n_scat_last_excursion_dist}
\end{figure}

To check the scalings in Eq.~(\ref{eq:N_scat_wing_approx}), we use MCRT to count the number of total scatterings, the number of scatterings in the core, and the number of scatterings in the excursion leading to escape for each photon. Each $\tau_0$ then gives a distribution, the mean of which follows the analytical scaling predictions. In the bottom panel of Fig.~\ref{fig:n_scat_wing_dist}, the distribution of the number of scatters in the wing (excluding the scatterings in the final excursion) is shown for various cloud optical depths. The points represent the mean of the distributions, and the arrows show the expected scaling prediction from one optical depth to the next. As predicted by the second term in Eq.~(\ref{eq:N_scat_wing_approx}), the number of wing scatterings scales linearly with $\tau_0$. The top panel shows the cumulative probability that a photon has less than $N_\text{scat,wing}$ scatterings prior to the final excursion. For lower optical depths, the CDF being $> 0$ at $N_\text{scat,wing} = 1$ indicates that the photons may escape in a single wing excursion without ever returning to the core. In the bottom panel of Fig.~\ref{fig:n_scat_last_excursion_dist}, the distribution of the number of scatterings in the excursion that led to escape is shown for various optical depths. The vertical line distinguishes the zero bin, e.g. the discrete probability that a photon does not scatter after making it to the wing for the final time. The arrows verify the expected scaling based on the previous mean and the analytical estimate from the first term in Eq.~(\ref{eq:N_scat_wing_approx}) as $(a\tau_0)^{2/3}$, especially between higher optical depths, where the diffusion assumptions are most valid. The top panel shows the cumulative distribution, the probability that a photon has $<N_\text{scat,last excursion}$ scatterings in the last wing excursion. At low optical depth, this is nonzero from the start since in optically thin clouds, escape in the core can occur. For higher optical depth, this verifies the $(a\tau_0)^{2/3}$ scaling for wing scatterings in the last excursion derived in \cite{LaoSmith2020} and re-derived later in Sec.~\ref{sec:radiation_properties}. This shows that photons escape into the wing and scatter several times prior to escape.

\begin{figure}
    \centering
    \includegraphics[width=\columnwidth]{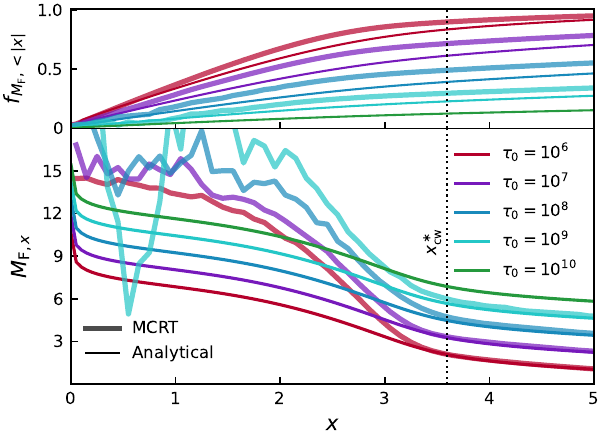}
    \caption{A comparison between the analytical spectral force multiplier (thin lines) and MCRT simulations (thick lines) as defined in Eq.~(\ref{eq:general_spectral_force_multiplier}). \textbf{Bottom Panel:} The analytical $M_{\text{F},x}$ from Eq.~(\ref{eq:point_spectral_force_multiplier}) evaluated exactly with numerical integration for various values of $\tau_0$ and for a temperature $T = 10^4 \, \K$ comparing to the binned contributions from MCRT simulations. The vertical dotted line shows the value of $x_\text{cw}^\ast$ for this temperature corresponding to where the qualitative behavior of the function changes. \textbf{Top Panel:} The cumulative fraction of $M_\text{F}$ from frequencies $<|x|$ ($f_{M_\text{F},<|x|}$) as defined in the text in Eq.~(\ref{eq:fractional_contribution}). The higher the optical depth, the more diffusion there is into the wing of the line leading to smaller contributions of the force multiplier from core scatterings.}
    \label{fig:M_F_x_core}
\end{figure}

\subsection{Spectral force multiplier}
The force multiplier describes the factor by which repeated resonant scatterings amplify radiation feedback relative to the single-scattering limit \citep{DijkstraLoeb2008, Kimm2018, LaoSmith2020, Nebrin2025}. \citet{Nebrin2025} demonstrates how this can be important even in the presence of damping mechanisms. In performing the MCRT simulations for \citet{Nebrin2025}, we found that there was some discrepancy between the analytical and numerical results, especially at intermediate optical depths. This can be attributed to the breakdown of the diffusion approximation. We find here that the diffusion approximation breaks down in the core specifically, while the analytical agreement with MCRT in the wing is very good. We define the spectral force multiplier by
\begin{equation} \label{eq:general_spectral_force_multiplier}
  M_{\text{F},x} = \frac{\text{d}M_\text{F}}{\text{d}x}
  \equiv \mathcal{L}^{-1}  \int k(\bm{r}) F \,\text{d}V 
  \approx -\frac{4\pi}{3\mathcal{L}}  \int \nabla J_x(\bm{r})\,\text{d}V \, .
\end{equation}
In the case of a point source in a static uniform density spherical cloud without dust \citep{LaoSmith2020}, Eq.~(\ref{eq:general_spectral_force_multiplier}) reduces to
\begin{equation} \label{eq:point_spectral_force_multiplier}
  M_{\text{F},x} = -\frac{\sqrt{8/3}}{\pi} \ln\left[\text{tanh}\left(\frac{\pi |\tilde{x}|}{2}\right)\right] \, .
\end{equation}
As described in Section~\ref{sec:intro}, core scatterings are not tracked when core-skipping acceleration schemes are used, leading to a missing momentum contribution that these photons would have every time they return to the core, since only one of the core scatterings is contributing when there could have been $\sim \tau_0$ scatterings that were skipped. To compute the force multiplier contributed by core scatterings, we notice that in the diffusion approximation in combination with the core approximation for $\tilde{x}$ leads to
\begin{equation}
    M_{\text{F},x}^\text{core} \approx  - \sqrt\frac{8}{3\pi^2} \ln\left(\sqrt{\frac{\pi^3}{24}} \,\frac{\text{erfi}(|x|)}{\tau_0}\right) \, ,
\end{equation}
and the contribution to $M_\text{F}$ in the wing is
\begin{equation}
    M_{\text{F},x}^\text{wing} \approx - \sqrt{\frac{8}{3\pi^2}} \ln\left[\tanh\left(-\sqrt{\frac{2 \pi^3}{27}} \frac{x^3}{a \tau_0}\right)\right] \, .
\end{equation}
The bottom panel of Fig.~\ref{fig:M_F_x_core} shows the spectral force multiplier $M_{\text{F},x}$ from Eq.~(\ref{eq:point_spectral_force_multiplier}) where $\tilde{x}$ is exactly evaluated with numerical integration. The temperature is fixed at $T = 10^4 \, \K$ and the different curves correspond to various optical depths. The vertical line shows the (constant) value of $x_\text{cw}^\ast$, which shows the qualitative change in the core-wing transition. The top panel shows the cumulative fraction of contribution to the force multiplier $f_{M_{\text{F},<|x|}}$ from frequencies $<|x|$. It is clear that for high $\tau_0$, wing scatterings dominate the force multiplier. However, there is a significant discrepancy for core frequencies, while the agreement in the wing is excellent. In general, analytical prediction significantly underestimates numerical results for core frequencies, resulting in the discrepancy in the total (integrated) force multiplier $M_\text{F}$ seen in previous papers, the result being more exaggerated for intermediate optical depths, as suggested by the cumulative fraction at $x_\text{cw}^\ast$ \citep{Nebrin2025,Smith2025}. In addition, the numerical results do not fully converge, especially for higher $\tau_0$. This can be caused by the lack of photons and $M_{\text{F},x}$ needing many photons ($\sim 10 \times \tau_0$ photons) to converge; however, the exorbitant cost of running MCRT simulations without core skipping is not always justifiable. Another reason is the momentum binning. The MCRT results calculate the net change in momentum by incoming frequency. The $R_\text{II}$ redistribution function shows that when the incoming frequency is in the wing, the outgoing frequency distribution is centered around the incoming frequency. If the incoming frequency is in the core, the outgoing frequency is effectively randomly redistributed into the core with a small chance of making a wing excursion \citep{Hummer1962, Dijkstra2014}. Hence, even if the binned $M_{\text{F},x}$ is noisy and remains so as the number of photons increases, the integrated value still converges (Kasiri et al., in prep).

The disagreement between the analytical and numerical solution is a fundamental breakdown in the diffusion approximation. To explore this, we show in Appendix~\ref{sec:rf_moments} that the Fokker--Planck approximation, described in Section~\ref{sec:intro}, breaks down for core frequencies since a term that is important in the core is ignored, which \citet{Rybicki1994} simply accepted as a necessary error to enforce photon conservation. However, photon conservation can be enforced if we expand the product of $R_\text{II} J$ as opposed to just $J$. The result, following \cite{Meiksin2006}, gives
\begin{align} \label{eq:fokker-planck-third-order}
    -k_xJ_x + &\int k_{x'} J_{x'} R_{x' \rightarrow x} \, \text{d}x' \approx
    \frac{1}{2}\frac{\partial}{\partial x}\left[k_x\frac{\partial J_x}{\partial x} + \frac{1}{3} \frac{\partial}{\partial x} \left(\frac{\partial^2 k_x}{\partial x^2} J_x\right)\right] \, ,
\end{align}
but even with this term and the numerical solution of the diffusion equation from initial explorations with \textsc{lydion-1d} (Nebrin et al., in prep), it is not obvious this improves the accuracy of $M_\text{F}$ calculations. A potential fix could be an even higher-order Fokker--Planck expansion, which is necessary in the core, since $|x - x'|/|x'|$ is potentially large, however physically the random-redistributive nature of the $R_\text{II}$ function in the core makes frequency diffusion inaccurate. An in-progress study of generalized frequency redistribution will explore these nuances.

\subsection{Uniform source solutions}
In this brief subsection, we derive the corresponding spectral distributions for the limiting case of a uniform source with constant density and no velocity structure where the emission is uniformly spread throughout the cloud. The general solutions for spherical geometry in this case are neatly presented in \cite{LaoSmith2020}. The spectral trapping time derived from Eq.~(\ref{eq:general_spectral_trapping_time}) and eq.~(99) in \cite{LaoSmith2020} is
\begin{equation}
    \frac{t_{\text{trap},x}}{t_\text{light}} = \frac{3 \sqrt6}{\pi^3} \text{Li}_3\left(e^{-\pi|\tilde{x}|}\right) \, ,
\end{equation}
where the polylogarithm is $\text{Li}_s(z) = \sum_{n=1}^\infty z^n/n^s$. Multiplying by $\tau_0 H(a,x)$ gives the spectral number of scatterings as
\begin{equation}
    N_{\text{scat},x} = \left[ \frac{3 \sqrt6}{\pi^3}H(a,x) \, \text{Li}_3 \left(e^{-\pi|\tilde{x}|}\right) \right] \tau_0 \, ,
\end{equation}
and finally the spectral force multiplier for a uniform source from Eq.~(\ref{eq:general_spectral_force_multiplier}) using eq.~(95) from \cite{LaoSmith2020} gives
\begin{align}
    M_{\text{F},x} = \frac{2\sqrt6}{\pi^3} \bigg[ \text{Li}_3 \Big(e^{-\pi|\tilde{x}|}\Big)  -  \text{Li}_3 \Big(-e^{-\pi|\tilde{x}|}\Big) \bigg] \, .
\end{align}
This analytical solution is shown in Fig.~\ref{fig:M_F_estimate} with the blue curves corresponding to the solutions using different calculation methods to get $M_{\text{f},x}$. These methods are explained in the following subsection, but the analytical solution works very well in the uniform case, as is expected from \cite{Nebrin2025} where the uniform source solutions matched the analytical solutions more neatly.

% \subsection{Accounting for core-skipping}
\subsection{Diffusion-based force estimators}
As mentioned above, numerous core scatterings slow down MCRT codes by several orders of magnitude. To achieve significant speed-ups, core scatterings are skipped by preferentially selecting interacting hydrogen atoms with enough velocity to move the photon out into the wing. Core-skipping algorithms come in multiple flavors, the simplest of which sets a fixed critical frequency $x_\text{crit}$. The photons inside the core, typically defined as $x_\text{cw}$ rather than $x_\text{cw}^\ast$ (see Fig.~\ref{fig:xcwast}), are then minimally pushed out to frequencies $x > x_\text{crit}$. In a more complex manner, $x_\text{crit}$ can be set dynamically by analyzing the minimum optical depth needed for escape. From the previous subsections, it is obvious that skipping these scatterings leads to errors in some internal radiation properties. We now discuss potential solutions to avoid needing high-photon, no core-skipping MCRT, which is too computationally burdensome in practice. Firstly, if only emergent radiation properties are needed, and core-skipping is safe to use, then we recommend using $x_\text{cw}^\ast$ as opposed to $x_\text{cw}$ as the boundary between core and wing frequencies, which the plot of $N_{\text{scat},x}$ (Fig.~\ref{fig:n_scat_x}) clearly shows is the boundary where the power-law scaling of the wing starts. 

If internal radiation properties are needed, then core skipping will give wrong results even if fully converged, since the core portion of the spectral distributions is almost completely lost. Core scatterings can account for anywhere from $5\% - 80\%$ of the total force multiplier (depending on optical depth) and account for virtually all scatterings. In the diffusion regime, path-based MCRT estimators for energy density and pressure converge faster than direct event-based MCRT momentum exchange, while also giving a force multiplier that is closer to no core skipping calculations than the analytical solution. The analytical estimate for the number of scatterings $\leq x_\text{cw}^\ast$ is very accurate, so we focus on strategies to accelerate the convergence of $M_\text{F}$. The most efficient, but perhaps resulting in the lowest accuracy correction, is to add $\int_{|x|<x_\text{cw}^\ast} M_{\text{F},x}^\text{core}$ to the force multiplier after completing a MCRT simulation with core skipping based on the analytical estimate. Either way, we suggest using the gradient of the energy density or the divergence of the pressure tensor integrated over the volume \citep{Roth2015}, as suggested by Eq.~(\ref{eq:M_F_def}). The energy density calculation converges much faster than computing the momentum transfers since energy density is computed with a path-based approach as opposed to a scattering-based one (see Kasiri et al., in prep). The pressure approach converges slower than the energy density gradient, but it is more fundamental to Fick's law in Eq.~(\ref{eq:Ficks_Law}) from which Eq.~(\ref{eq:M_F_def}) for the force multiplier follows, and allows for an anisotropic force in multi-dimensional geometries.

\begin{figure}
    \centering
    \includegraphics[width=\columnwidth]{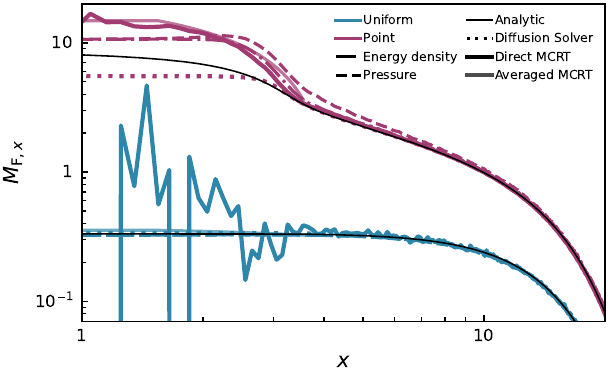}
    \caption{The spectral force multiplier $M_{\text{F},x}$ computed with various numerical techniques compared to our analytical solutions for both a uniform source and a point source using a constant optical depth $\tau_0 = 10^7$ and a constant temperature $T = 10^4 \, \K$. The numerical techniques include direct total change in momentum from MCRT, the gradient of the energy density and pressure, and from \textsc{Lydion-1d}, an upcoming Ly$\alpha$ diffusion solver.}
    \label{fig:M_F_estimate}
\end{figure}

In Fig.~\ref{fig:M_F_estimate} we show $M_{\text{F},x}$ computed using the different strategies described above for $\tau_0 = 10^7$ and $T = 10^4 \,\K$. To do this in MCRT, an internal spectrum tracker is used to find $u(\bm{r},x)$. We find that the analytical solution breaks down in the core and results in the lowest $M_\text{F}$ when integrated over all frequencies. The \textsc{lydion-1d} diffusion solver result for a point source is not converged since that would require very fine radial binning and hence it appears even lower, but the force multiplier agrees with the pressure calculation from MCRT. The \textsc{lydion-1d} result for the uniform source is converged. Calculating $M_\text{F}$ from the energy density has better agreement, but the error is still large in the core. The pressure estimator is even closer to the actual value of $M_\text{F}$, however, disagrees with the energy density estimator in the core and for frequencies near the core for the point source, which shows a better initial agreement with the direct momentum computation. For a uniform source, all strategies agree, but the direct momentum computation still gives the highest force multiplier. Note that the integrated $M_\text{F}$ value yet again is converged and the noise in the core is due to more random redistribution in the core so that many more photons are required for full convergence.

\begin{figure}
    \centering
    \includegraphics[width=\columnwidth]{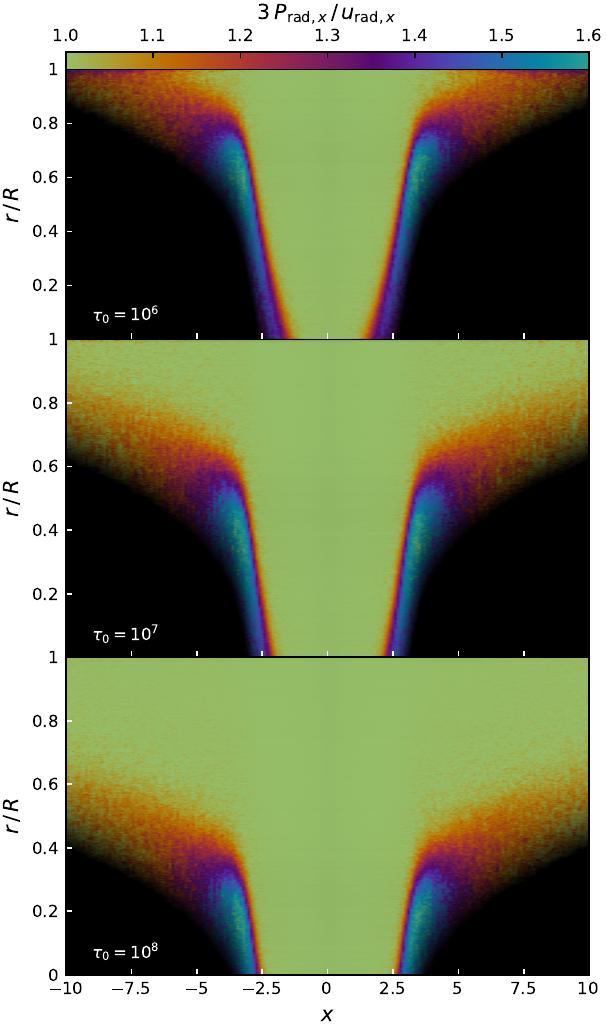}
    \caption{The Eddington Factor $3 \, P_{\text{rad},x}\,/\,u_{\text{rad},x}$ for optical depths $\tau_0 = 10^6$ (top panel), $\tau_0 = 10^7$ (middle panel), and $\tau_0 = 10^8$ (bottom panel) vs. normalized radius $r\,/\,R$ and frequency $x$. The temperature is fixed at $T = 10^4 \, \K$, and each simulation uses $10^5$ photons (enough for convergence). The images are placed on a black background and the opacity of each data cell is weighted by the normalized energy density at each radius. The expected value is $1$, corresponding to the lower limit of the color bar.}
    \label{fig:eddington_factor}
\end{figure}

The discrepancy between the energy density calculation and the pressure calculation in the point source calculations implies that the Eddington approximation ($P_\text{rad} = u_\text{rad}/3$) is not valid at certain frequencies and positions within the cloud. The pressure tensor is diagonalized under the isotropic radiation field approximation, and the eigenvalues should all be $P_\text{rad}(r) = u_\text{rad}(r)/3$ in spherical geometry. We plotted the frequency-dependent Eddington factor, defined as $P_\text{rad,x}(r)/u_\text{rad,x}(r)$, in Fig.~\ref{fig:eddington_factor} for a variety of $\tau_0$ values in the different panels and a spherical point source with $T = 10^4 \,\K$. The expected value is $3 \times \text{Eddington Factor} = 1$, which is the lower limit of the color bar. As also suggested in Fig.~\ref{fig:M_F_estimate} in the discrepancy between the energy density and pressure calculations, the Eddington factor is higher than $1/3$ near $\sim x_\text{cw}$, especially at lower radii. The more optically thick the cloud (or consequently the larger the radius), the more isotropic the radiation field becomes. This explains why the approximation is better near the edge of the cloud and at smaller radii for higher cloud optical depths. Overall, the Eddington approximation still holds in the limit $\tau_0 \rightarrow \infty$, however, for optical depths that are computationally affordable, there is a breakdown near the source, which may contribute to some error in the analytic solutions.

\section{Photon-based transport and escape statistics} \label{sec:radiation_properties}
In the previous section, we derived spectral distributions of internal radiation properties from the diffusion approximation to characterize the relative contribution of core scatterings and wing scatterings, and we compared our findings with MCRT simulations. The robustness of MCRT allows us to keep track of more information than is possible to track analytically. In this subsection, we present some statistics related to Ly$\alpha$ escape and scattering to understand the properties of Ly$\alpha$ radiation at different frequencies, enhancing our understanding of resonant-line transfer properties and escape, and complementing the results of Sec.~\ref{sec:spectral_distributions}. Ly$\alpha$ photons are traditionally thought to escape in a single longest excursion to the wing, where the optical depth drops to order unity, allowing the photon to diffuse in space. As a result, the higher the line center optical depth $\tau_0$, the higher the frequency must be for escape to become probable. However, escape to larger frequencies is rare, so a photon can undergo many small excursions, and it is unclear whether these can result in escape before the photon returns to the core. We compute the fraction of photons that escape a frequency, the expectation value of the total path length traversed, the total number of scatterings undergone, the radius reached, and the sum of the changes in frequency prior to a photon escaping a given frequency bin. Each expectation value is computed as a function of the max-frequency bin a photon has visited until the photon escapes the cloud. All simulation data come from the same set of no core-skipping MCRT simulations that fix the temperature at $T = 10^4 \, K$ while the number of photons was increased as high as modest computational resources allowed.

To explain the power-law scalings that will appear for wing frequencies, we review some prior scaling relations with the additional rigor of this paper. \cite{Osterbrock1962} derived the drift back to the core for wing frequencies as $-1/x$ and the RMS frequency displacement as $x$ resulting in $\sim x^2$ scatterings to return to the core. In Appendix~\ref{sec:return_to_core}, we derive this result in a more rigorous manner with a small correction and test it with simulations. In the wing, the photon travels a greater distance, and if it escapes, then standard random walk arguments give $N_\text{scat} \approx (R/\lambda_\text{mfp})^2$. Now in the wing, $\lambda_\text{mfp} = R/\tau_0 H(x) \approx \sqrt\pi Rx^2/a\tau_0$. Equating these two gives a characteristic escape frequency as $x_\text{esc} = (a\tau_0/\sqrt{\pi})^{1/3}$. From this we recover that for the escape excursion $N_\text{scat,last excursion} \propto (a\tau_0)^{2/3}$ which is consistent with the rigorous derivation from Eq.~\ref{eq:N_scat_wing_approx}. For a homogeneous medium, $\tau(r) = (r/R) \tau_0$, so $\tau_0$ can be replaced with $\tau(r)$ for inner rings of our spherical clouds of radius $r$.

\begin{figure}
    \centering
    \includegraphics[width=\columnwidth]{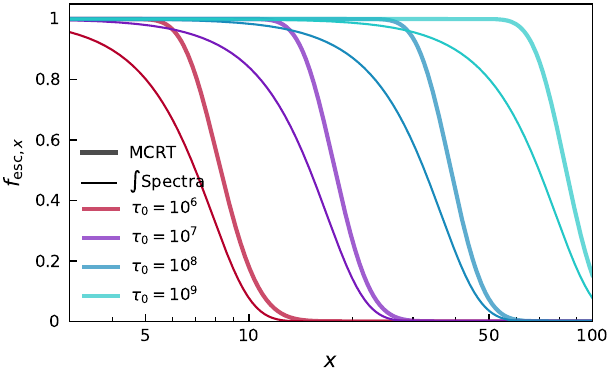}
    \caption{The faction of photons that escape to a frequency $\geq |x|$, $f_{\text{esc},x}$ for various optical depths with data from MCRT simulations. Normalizing the frequency axis by $(a\tau_0)^{1/3}$ results in nearly an identical function with small deviations from one $\tau_0$ to another. The integrated spectra from Eq.~(\ref{eq:integrated_spectra}) are below the numerical curves because they represent the emergent values allowing for drifts back to lower frequencies than the maximum excursion.}
    \label{fig:f_xesc}
\end{figure}
Fig.~\ref{fig:f_xesc} shows the fraction of photons that escape at a frequency $\geq |x|$ in thick lines and compares it with the complement of the integrated spectra using eq.~(88) from \citet{LaoSmith2020} (see also \citealt{Dijkstra2006}) in thin lines as a reference for various values of $\tau_0$. The integrated value of the spectra is
\begin{equation} \label{eq:integrated_spectra}
    1- \frac{\int_{-x}^x J(x) \, \dd x}{\int_{-\infty}^\infty J(x) \, \dd x} = 1 - \text{tanh}\left(\sqrt{\frac{\pi^3}{54}} \frac{x^3}{a\tau_0}\right) \, ,
\end{equation}
which is an estimate of the escape fraction. Note that the value of these two is not expected to agree since photons can escape to a high frequency and then drift back towards the core with subsequent wing scatterings that have a small preferential drift $-1/x$ \citep[see Appendix~\ref{sec:return_to_core}, and][]{Osterbrock1962}. Photons must surpass larger frequencies as $\tau_0$ increases, indicating a wider range of trapping even for frequencies further from resonance in the diffusion limit. The frequency needed for escape to be likely (e.g. when $f_{\text{esc},x} = 0.5$) scales as $(a\tau_0)^{1/3}$, as expected from the scaling arguments.

\begin{figure}
    \centering
    \includegraphics[width=\columnwidth]{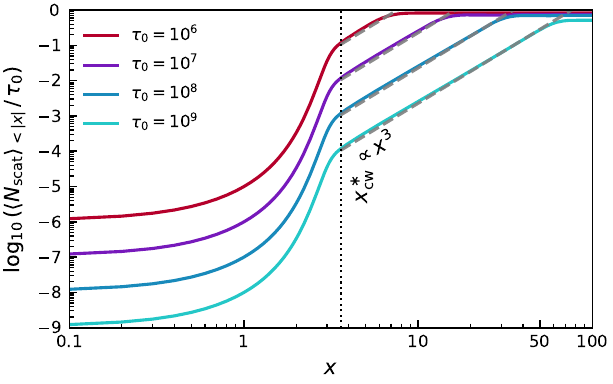}
    \caption{The average number of scatterings normalized by $\tau_0$ before a Ly$\alpha$ photon escapes the frequency $|x|$ for various optical depths with data from MCRT simulations. The dotted vertical line shows $x_\text{cw}^\ast$ from Eq.~(\ref{eq:xcwast_fit}).}
    \label{fig:n_scat_xesc}
\end{figure}

The average number of scatterings undergone before a photon escapes a frequency $|x|$ normalized by optical depth, as suggested by Eq.~(\ref{eq:N_scat_delta}), is plotted for different optical depths in Fig.~\ref{fig:n_scat_xesc}. The average number of scatterings to escape the core remains roughly the same as there is an order of magnitude difference between each curve in $\langle N_\text{scat} \rangle_{<|x|}/\tau_0$ at the frequency $|x| =x_\text{cw}^\ast$. The order of magnitude difference in $\tau_0$ shows the constancy. The power law scaling in the wing observed before the physical cutoff is $\langle N_\text{scat}\rangle_{<|x|} / \tau_0 \propto x^3$. This can be understood from the perspective of a first passage problem. The optical depth is proportional to the radius of the cloud due to the homogeneity of the cloud. In addition, we know that the characteristic number of scatterings to escape a cloud of optical depth $\tau(r)$ is just $\tau(r)$, which characterizes the number of scatterings needed to escape the core at such an optical depth \citep{LaoSmith2020}. From the scaling arguments, a wing frequency of $x_\text{esc} \propto [a\tau(r)]^{1/3}$ is needed for escape, so $\langle N_\text{scat}\rangle_{<|x|} \propto \tau(r) \propto x^3$. This explains the scaling in the wing. After $\sim \tau_0$ scatterings, the photons eventually gets a lucky excursion and escape the cloud, resulting in the physical cutoff. This asymptotic flattening corresponds to the average number of scatterings to escape the cloud, which scales as $\tau_0$, but not exactly by an order of magnitude, since the diffusion approximation becomes more accurate as $\tau_0 \rightarrow \infty$, and we find a better agreement with Eq.~(\ref{eq:N_scat_delta}). 

\begin{figure}
    \centering
    \includegraphics[width=\columnwidth]{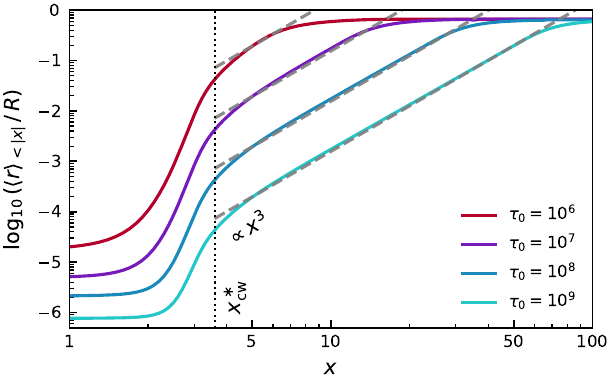}
    \caption{The average radius in units of cloud radius before a Ly$\alpha$ photon escapes the frequency $|x|$ for various optical depths with data from MCRT simulations. The dotted vertical line shows $x_\text{cw}^\ast$ from Eq.~(\ref{eq:xcwast_fit}).}
    \label{fig:r_xesc}
\end{figure}

Fig.~\ref{fig:r_xesc} shows the average radius reached before a photon escapes a frequency $|x|$, which we call $\langle r \rangle_{<|x|}$, normalized by the cloud radius. The average radius to escape $x_\text{cw}^\ast$ is $\propto \tau_0^{-1}$ since photons in optically thicker environments have a lower mean free path across frequencies, making escape from the core more difficult until a sufficiently large frequency is reached. The power law scaling in the wing is as in the case of the average number of scatterings $\langle r \rangle_{<|x|}/ R \propto x^3$. This scaling is explained similarly by the mean free path argument. The asymptotic value of $\sim 0.7$ approached by all values of $\tau_0$ characterizes the average radius reached before a photon escaped a cloud without reaching a higher frequency. This does not mean that the photon will not scatter at a radius $ \geq 0.7 R$, but, on average, the photon will not drift to a higher frequency beyond that radius.

\begin{figure}
    \centering
    \includegraphics[width=\columnwidth]{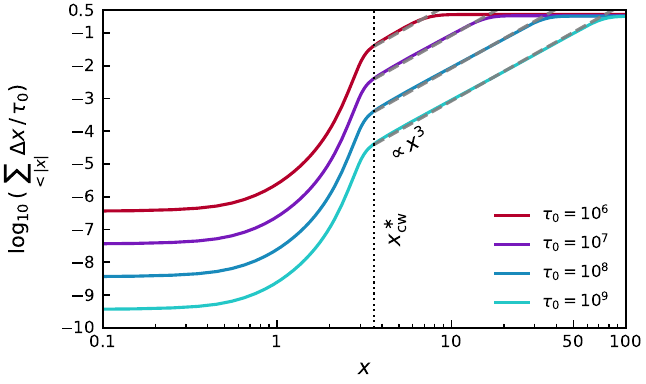}
    \caption{The average total change in frequency before a Ly$\alpha$ photon escapes the frequency $|x|$ for various optical depths with data from MCRT simulations. The dotted vertical line shows $x_\text{cw}^\ast$ from Eq.~(\ref{eq:xcwast_fit}).}
    \label{fig:dx_tot_xesc}
\end{figure}

Fig.~\ref{fig:dx_tot_xesc} shows the cumulative sum of all frequency changes before escaping a frequency $|x|$, that is, $\sum_{<|x|} \Delta x /\tau_0$. Note that this is binned by the max frequency bin reached, so once a photon has an absolute frequency $> |x|$, changes in frequency do not contribute to the sum even if the frequency goes below $|x|$ subsequently. We normalize this by $\tau_0$ since, on average, the frequency displacement between scatterings is $\Delta x \approx 1$ \citep[see Appendix~\ref{sec:rf_moments} for the first moment of the redistribution function, and][]{Osterbrock1962}, so that $\sum \Delta x \sim N_\text{scat} \propto \tau_0$. As in the case of $\langle N_\text{scat} \rangle$, the total frequency displacement before leaving the core is roughly constant across optical depths. The wings scale with frequency as $\sum_{<|x|} \Delta x / \tau_0 \propto x^3$, and this scaling is explained by the same argument as for the average number of scatterings since $\sum \Delta x \sim \tau_0$. The similar asymptotic value $\Delta x_\text{tot} \approx 0.4 \tau_0$ indicates that $\sum_{<|x|} \Delta x$ scales with $\tau_0$.

\begin{figure}
    \centering
    \includegraphics[width=\columnwidth]{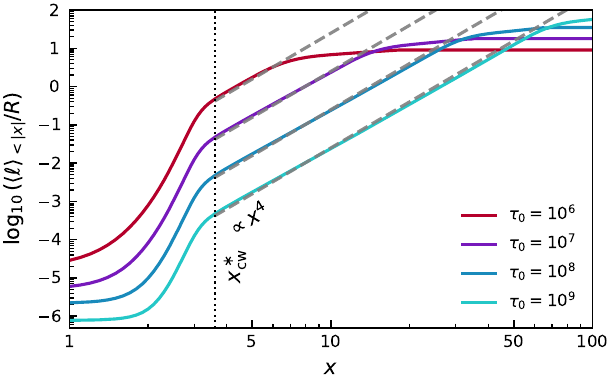}
    \caption{The average total path length traversed by a Ly$\alpha$ photon in units of cloud radius before escaping the frequency $|x|$ for various optical depths with data from MCRT simulations. The dotted vertical line shows $x_\text{cw}^\ast$ from Eq.~(\ref{eq:xcwast_fit}). There is a strong bias to experience most of the transport near the expected critical escape frequency $x_\text{esc} \approx (a\tau_0)^{1/3}$.}
    \label{fig:l_tot_xesc}
\end{figure}

Fig.~\ref{fig:l_tot_xesc} is the average path length traversed before reaching a frequency $|x|$, normalized by the cloud radius for various optical depths. As the optical depth increases, the average path length to escape the core decreases since core photons are extremely optically thick, so the distance between scatterings is very small, and photons cannot diffuse in space until they undergo a random wing excursion. Outside of the core, the average path length scales as a power law in frequency $\langle \ell \rangle_{<|x|}/R\propto x^4$. The extra factor of x arises from the simple fact that $\langle \ell \rangle_{<|x|} \approx N_\text{scat, last excursion} \lambda_\text{mfp}$. In the wing for the excursion that leads to escape, we know $N_\text{scat} \propto x_\text{esc}^2$ and $\lambda_\text{mfp} \propto x_\text{esc}^2$ in the wings, so $\langle \ell \rangle_{<|x|}/R\propto x^4$. The curves flatten asymptotically once most photons have escaped the cloud, which couples physical space to frequency space. The frequency at which the curves depart from the power-law scaling is comparable to the frequency that only $\sim 50 \%$ of photons reach, which can be seen in Fig.~\ref{fig:f_xesc}. These statistics provide strong evidence that $x_\text{cw}^\ast$ is the singular frequency that represents the boundary between the core and the wing, as the characteristic power law scaling in the wings begins to be valid around this frequency across all statistics. Additionally, we gain insight into the coupling of the diffusion process in both frequency and physical space and how the spatial boundary conditions cause asymptotic behavior for many statistics.

\section{Anomalous spatial but classical spectral diffusion} \label{sec:anomalous_diffusion}
In this section, we explore the distribution of jump lengths and the overall change in frequency between successive scatterings, suggesting anomalous diffusion due to the fat-tailed wings of their probability density functions. In the previous section, all figures demonstrated diffusion in both frequency and physical space, which are tightly intertwined depending on whether the photon was in the wing or the core, but the analytical treatment of optically thick radiative transfer as a double diffusion process is not valid across all frequencies because of the breakdown of both Fick's law and Fokker--Planck for core frequencies (see Fig.~\ref{fig:M_F_x_core}). This raises the question of what the accurate model is. Recently, anomalous diffusion was pointed out as a potential model for cosmic rays, and it was pointed out that in resonant-line radiative transfer the Lorentzian tails of $H(x) \propto x^{-2}$ in the wings can give rise to `Levy-like' behavior \citep{LiangOh2025, AlmadaMonter2025}. If certain distributions are fat-tailed, then this can help explain the breakdown of the diffusion approximation and suggests a different treatment of diffusion, specifically when the incoming or outgoing frequency is in the core of the line $|x| \leq x_\text{cw}^\ast$. The alternative, to be explored in future work, is fractional diffusion models that can accurately capture the non-local effects of anomalous diffusion \citep{Metzler2000}. We also numerically compute frequency statistics on escaped photons and present the probability that a photon that escapes has some frequency by calculating the minimum and maximum frequencies reached in the wing excursion leading to escape, which leads to insights into the behavior of frequency diffusion around core frequencies. 

\begin{figure}
    \centering
    \includegraphics[width=\columnwidth]{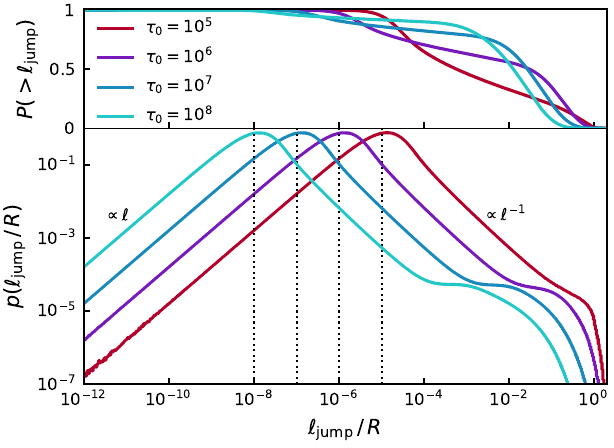}
    \caption{Statistics on photon jump lengths $\ell_\text{jump}$ between successive scatterings normalized by the cloud radius $R$ for various values of $\tau_0$ all for a fixed temperature of $T = 10^4 \,\K$. \textbf{Bottom Panel}: The probability density function $p(\ell_\text{jump}/R)$. The vertical dotted lines show $1/\tau_0$, which is roughly the predicted mean free path. The tails of the distribution are power laws. \textbf{Top Panel}: The cumulative distribution function $1 - P(<\ell_\text{jump}) = P(>\ell_\text{jump})$ that gives the probability a photon had a jump $>$ $\ell_\text{jump}$.}
    \label{fig:jump_distribution}
\end{figure}

In the bottom panel of Fig.~\ref{fig:jump_distribution}, the distribution of path lengths between scattering events, normalized by the cloud radius, is shown for various optical depths. The vertical dotted lines are at values $1/\tau_0$, as Eq.~(\ref{eq:N_scat_delta}) suggests as $\sim$ the mean free path $\langle \ell_\text{jump} \rangle$ and is very near the peak of each distribution. Both tails of this distribution follow a linear power law ($\propto \ell_\text{jump}$ on the left wing and $\propto \ell_\text{jump}^{-1}$ on the right wing), which continues until Monte Carlo sampling limits the distribution on the lower end, and the radius of the cloud coupled with the optical depth limits the upper end. Note that the maximum jump length is $2 R$. As is the argument in the cosmic-ray community, this physical limit cuts off the fat tail of the distribution, limiting the effect of anomalous diffusion in practice \citep{LiangOh2025}. The top panel shows the probability that a photon had a jump $> \ell_\text{jump}$ normalized by radius. We note that as the optical depth increases there is an additional transition where spatial transport of wing photons becomes non-anomalous since the mean-free-path at the critical escape frequency becomes much smaller than the cloud radius.

\begin{figure}
    \centering
    \includegraphics[width=\columnwidth]{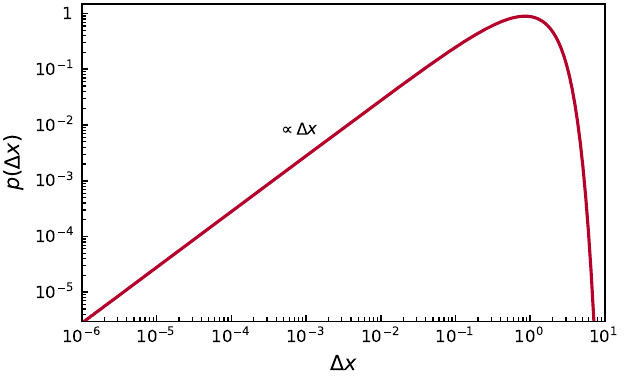}
    \caption{The probability distribution of frequency shifts between successive scatterings $p(\Delta x)$. This is plotted for a fixed optical depth value of $\tau_0 = 10^7$ and $T=10^4$, however the distribution is independent of $\tau_0$ and $T$. The lower end follows a power law scaling but the distribution has an exponential cutoff.}
    \label{fig:freq_jump_distribution}
\end{figure}

In Fig.~\ref{fig:freq_jump_distribution}, the distribution of changes in frequency between scattering events is shown for various optical depths. The distribution has linear scaling as $\Delta x \rightarrow 0$, but decreases exponentially for larger jumps. Linear scaling arises from the coherent nature of core scatterings (the flat profile of $R(x,x')$ when $x, x' \lesssim x_\text{cw}$), which integrates to give linear scaling. Once $\Delta x \sim x_\text{cw}$ the scattering couples to space and causes the exponential cutoff since a photon that undergoes a large change in frequency is likely to escape from the cloud with successive wing scatterings of order unity. The exponential decay in the redistribution function for core-wing transitions also contributes to this cutoff. The shape is independent of the optical depth as increasing the optical depth simply increases the number of interactions, and thus has no impact on individual scattering statistics. Changing the temperature changes the width of the Maxwellian velocity, so the frequency shifts in physical units $\Delta \nu$ are different, but the temperature dependence is encoded in $\Delta x = \Delta\nu/\Delta \nu_\text{D}$ through the Doppler width $\Delta \nu_\text{D} \propto T^{1/2}$ and thus in dimensionless units there is no temperature dependence. This was verified with simulations across different optical depths and temperatures. This distribution can be understood as the weighting of the redistribution function with a given frequency shift $\Delta x$ with the average internal spectrum $\langle J_x \rangle \equiv \int J(r,x) \,\dd V / \int \dd V$:
\begin{equation}
    p(\Delta x) = \frac{ \int_{-\infty}^{\infty} \langle J_{x'}\rangle R_\text{II}(x' +\Delta x, x') \,\dd x'}{\int_{-\infty}^{\infty} \langle J_{x'}\rangle \varphi(x') \,\dd x' } \, ,
\end{equation}
% \begin{align}
%     p(\Delta x) &= \frac{\int_V \int_{-\infty}^{\infty} J(r,x') R_\text{II}(x' +\Delta x, x') \,\dd x' \dd V}{\int_V \int_{-\infty}^{\infty} J(r,x') \varphi(x') \,\dd x' \dd V} \notag \\
%     &= \frac{ \int_{-\infty}^{\infty} \langle J_{x'}\rangle R_\text{II}(x' +\Delta x, x') \,\dd x'}{\int_{-\infty}^{\infty} \langle J_{x'}\rangle \varphi(x') \,\dd x' } \, ,
% \end{align}
where, to alleviate confusion, we define $R_{\text{II}}(x,x') = \varphi(x') R_{x' \rightarrow x}$ as the joint probability, normalized to unity after integrating over both incoming and outgoing frequencies, that a photon is absorbed at frequency $x'$ and remitted at frequency $x$ assuming that the ground state has no natural line width (the appropriate assumption for Ly$\alpha$, coined by \citet{Hummer1962} as type II redistribution) and $\langle J_x \rangle$ is the same as preceding Eq.~\ref{eq:spherical_point_spectral_t_trap}. 
%A similar integral is discussed in Appendix~\ref{sec:weighted-FP} to find the error of the Fokker-Planck approximation term in the RT equation using the analytical solution. 
The peak of $p(\Delta x)$ is $\sim 1$, changing frequency by one Doppler width as expected from the second moment of the redistribution function given in Appendix~\ref{sec:return_to_core}.

\begin{figure}
    \centering
    \includegraphics[width=\columnwidth]{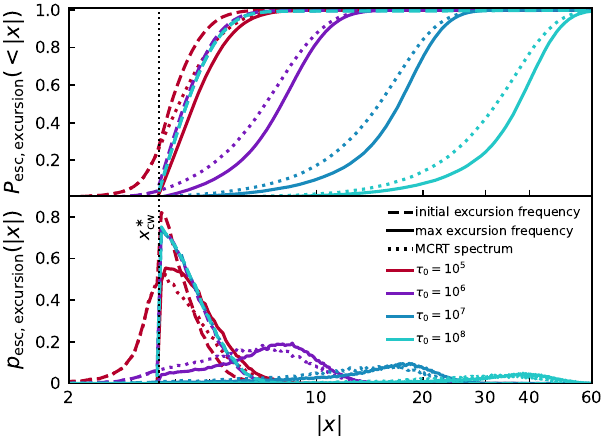}
    \caption{Statistics on photon wing excursions that lead to escape across several optical depths for a fixed $T = 10^4 \,\K$. The vertical dotted line denotes the value of $x_\text{cw}^\ast$. The dashed lines show the probability density that an initial excursion frequency leads to escape. The solid lines show the probability density of the max frequency reached during the wing excursion that leads to escape. Finally, the normalized spectra obtained from the MCRT simulations is plotted to show the shift from max excursion frequency to escape frequency.}
    \label{fig:esc_excursion}
\end{figure}

Fig.~\ref{fig:esc_excursion} shows statistics on wing excursions leading to photon escape for various optical depths. We track the initial frequency of each excursion, defined as the frequency the photon has after leaving the core ($x_\text{init}$), the maximum frequency reached during that excursion ($x_\text{max}$), and whether the photon escapes as a result of that excursion. If a photon escapes the cloud while $|x_\text{esc}| < x_\text{cw}^\ast$, then we count $x_\text{init} = x_\text{esc}$ since it never had an excursion that led to escape for us to track $x_\text{max}$. In the bottom panel, the dotted lines show the probability density that a photon that escapes has an initial excursion frequency $x_\text{init}$. Aside from the medium optical depth case ($\tau_0 = 10^5$), where diffusion in the wing does not hold even near the core, this distribution is universal outside of $x_\text{cw}^\ast$. The solid lines show the probability density that a photon reaches some $x_\text{max}$ during the excursion that leads to escape. We also plot the normalized emergent spectra obtained from the MCRT simulations $J(x)/\int J(x)\,\dd x$ and see that photons tend to drift back towards the core after reaching their maximum frequency before finally escaping. The top panel shows the cumulative probability that a photon had a frequency $<|x|$ as its maximum or initial escape frequency, as well as the cumulative spectrum, which is the complement of Eq.~(\ref{eq:integrated_spectra}). The drift in distributions from the initial excursion frequency distribution, to the max excursion frequency distribution, and finally the emergent spectra shows that photons that escape are mostly those that escape from the core with a frequency close to $x_\text{cw}^\ast$, rather than a large jump, and diffused out to a large enough frequency in several lucky wing scatterings that pushed the photons further out into the wing before returning diffusely prior to escape. The cumulative distribution reaching $1$ for $x\sim10$ further shows that most excursions have a small initial wing frequency before moving further into the wing. When the optical depth is medium, the distributions do not shift as much as even small escapes from the core render the opacity small enough to escape the cloud.

The statistics on excursions combined with the frequency change distribution are evidence that the standard diffusion picture holds outside the core boundary, but even near $x_\text{cw}^\ast$, as it is too rare to have large jumps either from the core out into the far wing or jumps from the far wing back into the core. The jump probability distribution, however, is fat-tailed on both ends indicating that even if RT is diffusive in frequency space, there can still be rare Levy flight behavior in physical space for individual photons.

\section{Summary and discussion}
\label{sec:summary}

In this paper, we have performed a detailed analysis of the frequency-dependent structure of Ly$\alpha$ radiative transfer in static, homogeneous spherical clouds. We focused on isolating core and wing contributions to internal radiation properties and understanding when diffusion-based analytic approximations remain reliable and where they lose accuracy. We summarize our main results from benchmarking high optical depth diffusion solutions with converged MCRT calculations without core-skipping:
\begin{enumerate}
  \item \textit{A physically motivated core--wing transition.}
  We introduced a new definition of the core--wing transition frequency, $x_\text{cw}^\ast$, based on the integrated diffusion variable $\tilde{x}$ rather than the local behavior of the Voigt profile alone, as was done for $x_\text{cw}$. This definition more cleanly delineates the change in transport regime between nearly coherent, non-diffusive core scatterings and genuine frequency diffusion in the wings. In practice, we find that using $|x| < x_\text{cw}^\ast$ as the trigger for core-skipping in MCRT codes is better aligned with the actual onset of wing-like behavior than the usual choice based on $H(a,x)$, and should therefore be preferred for use with core-skipping whenever only emergent properties are desired.

  \item \textit{Spectral trapping times and number of scatterings.}
  Using the diffusion approximation, we derived spectral distributions for the trapping time $t_{\text{trap},x}$ and the scattering rate $N_{\text{scat},x}$ for point and uniform sources. These analytic spectra agree extremely well with our no core-skipping MCRT results over the full frequency range once the $a\tau_0 \gg 10^3$ diffusion criterion is satisfied. Core photons contribute a nearly constant baseline independent of temperature and $\tau_0$ to $t_\text{trap}$, while the wings dominate the growth of $t_\text{trap}$ with optical depth. In contrast, the total number of scatterings $N_\text{scat}$ is almost entirely set by core interactions and scales $\propto \tau_0$ with only a small wing correction. We also rigorously derived and numerically confirmed the scaling $N_{\text{scat, last excursion}} \propto (a\tau_0)^{2/3}$ for the final wing excursion and $N_{\text{scat, wing}} \propto a\tau_0$ for the cumulative scaling for the number of wing scatterings that eventually return to the core.

  \item \textit{Limitations of the Diffusion approximation for the force multiplier.}
  We constructed analytic expressions for the spectral force multiplier $M_{\text{F},x}$ and compared them to MCRT estimates based on the total momentum exchange, the gradient of the energy density, and the gradient of the radiation pressure. In the wings, the diffusion solutions reproduce the numerical $M_{\text{F},x}$ remarkably well. However, in the core, diffusion theory systematically underestimates the true force. In the integrated results from \citet{Nebrin2025}, this is more noticeable for intermediate optical depths where the total fraction of force imparted by core scatterings is greater than for higher optical depths where wing scatterings dominate. We show that this discrepancy arises from a breakdown of the standard Fokker--Planck approximation and, more fundamentally, from the fact that scattering in the core is only weakly diffusive in frequency. On the numerical side, we find that MCRT force estimates converge slowly, with the required number of photon packets increasing at higher optical depths as a result of more scattering with near-isotropic cancellations. In practice, force estimators based on $\boldsymbol{\nabla} u_\text{rad}$ or $\boldsymbol{\nabla} \boldsymbol{\cdot} \mathsf{P}_\text{rad}$ converge substantially faster than direct momentum exchange and are therefore strongly recommended for precise Ly$\alpha$ radiation pressure feedback calculations. For a uniform source, we find good agreement between the momentum, pressure, and energy density calculations of spectral force multiplier. However, for point sources, there are still order unity relative differences at intermediate optical depths compared to converged MCRT results.

  \item \textit{Eddington and Fokker--Planck approximations.}
  By comparing energy density and pressure-based force estimators, we quantified where the Eddington approximation $P_\text{rad,x} = u_{\text{rad},x}/3$ fails inside the cloud. For point sources, the Eddington factor deviates most strongly from $1/3$ near the core--wing transition and at smaller radii, particularly for moderate $\tau_0$. These deviations decrease as $\tau_0$ increases and the radiation field becomes more isotropic and wing diffusion dominated. We confirm that the second-order Fokker--Planck approximation is strictly valid only in the wings, whereas in the core the redistribution kernel is broad. In most cases the Fokker--Planck and Eddington approximations can be safely applied.

  \item \textit{Escape channels and coupled spatial--spectral diffusion.}
  Using no core-skipping MCRT, we characterized how photon frequencies and spatial positions couple. The probability that a photon ever reaches a given frequency, the average number of scatterings, total path length traversed, and radius attained before first crossing $|x|$ all exhibit clear power-law scalings in the wings once $|x| \gtrsim x_\text{cw}^\ast$. In particular, we find $\langle N_\text{scat} \rangle_{<|x|} \propto x^3$, $\langle r \rangle_{<|x|}/R \propto x^3$, and $\langle \ell_\text{tot} \rangle_{<|x|}/R \propto x^4$, up to a cutoff set by the finite cloud size. The primary escape mechanism in optically thick clouds is not in extraordinarily large jump directly from the core to $x \sim (a\tau_0)^{1/3}$, but rather a sequence of failed and lucky wing excursions. Photons typically leave the core in modest steps just beyond $x_\text{cw}^\ast$, undergo multiple scatterings in the wings, and return to the core or, more rarely, diffuse out to $|x| \sim (a\tau_0)^{1/3}$ and escape.

  \item \textit{Anomalous diffusion and fat-tailed statistics.}
  Finally, we studied the statistics of spatial and frequency jumps between successive scatterings. The jump length distribution exhibits broad, power-law wings that are truncated only by the finite cloud radius, while the frequency shift distribution peaks at $|\Delta x| \sim \mathcal{O}(1)$ and shows a linear scaling for small $|\Delta x|$ followed by an exponential cutoff. Escape-excursion statistics indicate that frequency space is well-described by normal diffusion, but individual photons can undergo rare, long spatial jumps reminiscent of Lévy flights with power-law tails, however the radiative trasnfer equation does not immediately lend itself to fractional diffusion treatment because of the exponential falloff in the redistribution function for core--wing transitions. These fat-tailed statistics help explain why diffusion-based analytic treatments break down in the core and suggest that fractional or non-local diffusion formalisms may be required to accurately describe resonant-line transport in that regime.
\end{enumerate}

Taken together, these results provide a coherent framework for understanding where analytic diffusion treatments of Ly$\alpha$ radiative transfer are reliable and where full MCRT (or hybrid) methods are indispensable. For emergent spectra alone, core-skipping remains an efficient and accurate strategy, especially if the threshold is tied to the physically motivated $x_\text{cw}^\ast$. However, for internal quantities that control Ly$\alpha$ radiation pressure feedback, core scatterings are essential and diffusion-based estimates of the force multiplier systematically miss the core contribution, which are also missed with core-skipping. In such cases, analytic corrections, energy-density- or pressure-based force estimators, and diffusion solvers like \textsc{lydion-1d} (Nebrin et al., in prep.) can be combined to restore much of the missing physics at a fraction of the cost of brute-force MCRT.

Our analysis has focused on Ly$\alpha$ transport within idealized, static, dust-free, homogeneous spherical clouds. Future work could extend these methods to more realistic astrophysical environments, including dust absorption and destruction, inhomogeneous and clumpy media, and dynamical backgrounds with velocity gradients and time dependence. Incorporating anomalous-diffusion-inspired closures, or alternative treatments of the radiative transfer equation by splitting the scattering kernel for wing and core frequencies, as well as carefully validated core-skipping and diffusion schemes, into radiation-hydrodynamic simulations will be particularly important for quantifying Ly$\alpha$ radiation pressure as a pre-supernova feedback channel in high-redshift galaxies.

% \section{Remove this eventually}
% \begin{enumerate}
%   \item If only emergent radiation properties are needed, we recommend using $|x| < x_\text{cw}^\ast$ to trigger core-skipping instead of the traditional $x_\text{cw}$ core-to-wing transition frequency.
%   \item Excellent accuracy for analytic solutions for the spectral trapping time $t_{\text{trap},x}$ and number of scatterings $N_{\text{scat},x}$ both in the core and the wing. However, the analytic spectral force multiplier $M_{\text{F},x}$ underestimates the true value in the core, particularly at lower optical depths. MCRT solutions converge slowly, generally requiring $N_\text{photons} \gtrsim N_\text{scat} \sim \tau_0$. (Or is this $\sqrt{N_\text{scat}}$?)
%   \item Mention Eddington and Fokker--Planck approximations.
%   \item The primary escape mechanism in homogeneous clouds is...Need to conclude whether it is (1) rare excursions that already get you within a factor of $X$ of $x_\text{esc} \sim a\tau_0^{1/3}$ after many failed attempts, or (2) lucky diffusion that happens to move you further into the wing each time, or (3) both. (Can say as opposed to ... for clarity.)
% \end{enumerate}

\section*{Acknowledgements}
% We thank the referee for constructive comments and suggestions which have improved the quality of this work.
We thank Ethan Stace, Hyunbae Park, and Daniele Manzoni and the broader Ly$\alpha$ community for fruitful discussions. AS acknowledges support through HST AR-17859, HST AR-17559, and JWST AR-08709.

%%%%%%%%%%%%%%%%%%%%%%%%%%%%%%%%%%%%%%%%%%%%%%%%%%
% \section*{Data Availability}
% The data underlying this paper will be shared on reasonable request to the corresponding author.



\bibliographystyle{mnras}
\bibliography{biblio}

@ARTICLE{Partridge1967,
       author = {{Partridge}, R.~B. and {Peebles}, P.~J.~E.},
        title = "{Are Young Galaxies Visible?}",
      journal = {\apj},
         year = "1967",
        month = "Mar",
       volume = {147},
        pages = {868},
          doi = {10.1086/149079},
       adsurl = {https://ui.adsabs.harvard.edu/abs/1967ApJ...147..868P}
}

@ARTICLE{Dijkstra2014,
       author = {{Dijkstra}, Mark},
        title = "{Ly{\ensuremath{\alpha}} Emitting Galaxies as a Probe of Reionisation}",
      journal = {\pasa},
         year = "2014",
        month = "Oct",
       volume = {31},
          eid = {e040},
        pages = {e040},
          doi = {10.1017/pasa.2014.33},
archivePrefix = {arXiv},
       eprint = {1406.7292},
 primaryClass = {astro-ph.CO},
       adsurl = {https://ui.adsabs.harvard.edu/abs/2014PASA...31...40D}
}

@ARTICLE{Harrington1973,
       author = {{Harrington}, J. Patrick},
        title = "{The scattering of resonance-line radiation in the limit of large optical depth}",
      journal = {\mnras},
         year = "1973",
        month = "Jan",
       volume = {162},
        pages = {43},
          doi = {10.1093/mnras/162.1.43},
       adsurl = {https://ui.adsabs.harvard.edu/abs/1973MNRAS.162...43H}
}

@ARTICLE{Neufeld1990,
       author = {{Neufeld}, David A.},
        title = "{The Transfer of Resonance-Line Radiation in Static Astrophysical Media}",
      journal = {\apj},
         year = "1990",
        month = "Feb",
       volume = {350},
        pages = {216},
          doi = {10.1086/168375},
       adsurl = {https://ui.adsabs.harvard.edu/abs/1990ApJ...350..216N}
}

@ARTICLE{LoebRybicki1999,
       author = {{Loeb}, Abraham and {Rybicki}, George B.},
        title = "{Scattered Ly{\ensuremath{\alpha}} Radiation around Sources before Cosmological Reionization}",
      journal = {\apj},
         year = "1999",
        month = "Oct",
       volume = {524},
       number = {2},
        pages = {527-535},
          doi = {10.1086/307844},
archivePrefix = {arXiv},
       eprint = {astro-ph/9902180},
 primaryClass = {astro-ph},
       adsurl = {https://ui.adsabs.harvard.edu/abs/1999ApJ...524..527L}
}

@ARTICLE{Dijkstra2006,
       author = {{Dijkstra}, Mark and {Haiman}, Zolt{\'a}n and {Spaans}, Marco},
        title = "{Ly{\ensuremath{\alpha}} Radiation from Collapsing Protogalaxies. I. Characteristics of the Emergent Spectrum}",
      journal = {\apj},
         year = "2006",
        month = "Sep",
       volume = {649},
       number = {1},
        pages = {14-36},
          doi = {10.1086/506243},
archivePrefix = {arXiv},
       eprint = {astro-ph/0510407},
 primaryClass = {astro-ph},
       adsurl = {https://ui.adsabs.harvard.edu/abs/2006ApJ...649...14D}
}

@article{Ahn2002,
author = {Ahn, Sang-Hyeon and Lee, Hee-Won and Lee, Hyung Mok},
doi = {10.1086/338497},
issn = {0004-637X},
journal = {ApJ},
month = mar,
pages = {922--930},
title = {{Ly$\alpha$ Line Formation in Starbursting Galaxies. II. Extremely Thick, Dustless, and Static H I Media}},
url = {http://adsabs.harvard.edu/abs/2002ApJ...567..922A http://adsabs.harvard.edu/cgi-bin/nph-data\_query?bibcode=2002ApJ...567..922A\&link\_type=ARTICLE},
volume = {567},
year = {2002}
}

@article{Zheng2002,
author = {Zheng, Zheng and Miralda-Escud\'{e}, Jordi},
doi = {10.1086/342400},
issn = {0004-637X},
journal = {ApJ},
month = oct,
pages = {33--42},
title = {{Monte Carlo Simulation of Ly$\alpha$ Scattering and Application to Damped Ly$\alpha$ Systems}},
url = {http://adsabs.harvard.edu/abs/2002ApJ...578...33Z http://adsabs.harvard.edu/cgi-bin/nph-data\_query?bibcode=2002ApJ...578...33Z\&link\_type=ARTICLE},
volume = {578},
year = {2002}
}

@article{Osterbrock1962,
author = {Osterbrock, Donald E.},
doi = {10.1086/147258},
issn = {0004-637X},
journal = {ApJ},
month = {jan},
pages = {195},
title = {{The Escape of Resonance-Line Radiation from an Optically Thick Nebula.}},
url = {http://adsabs.harvard.edu/abs/1962ApJ...135..195O http://adsabs.harvard.edu/cgi-bin/nph-data{\_}query?bibcode=1962ApJ...135..195O{\&}link{\_}type=ARTICLE},
volume = {135},
year = {1962}
}

@ARTICLE{Adams1972,
       author = {{Adams}, Thomas F.},
        title = "{The Escape of Resonance-Line Radiation from Extremely Opaque Media}",
      journal = {\apj},
         year = 1972,
        month = jun,
       volume = {174},
        pages = {439},
          doi = {10.1086/151503},
       adsurl = {https://ui.adsabs.harvard.edu/abs/1972ApJ...174..439A}
}

@article{Adams1975,
author = {Adams, Thomas F.},
doi = {10.1086/153891},
issn = {0004-637X},
journal = {ApJ},
month = {oct},
pages = {350--351},
title = {{The Mean Photon Path Length in Extremely Opaque Media}},
url = {http://adsabs.harvard.edu/abs/1975ApJ...201..350A http://adsabs.harvard.edu/cgi-bin/nph-data{\_}query?bibcode=1975ApJ...201..350A{\&}link{\_}type=ARTICLE},
volume = {201},
year = {1975}
}

@ARTICLE{HansenOh2006,
       author = {{Hansen}, Matthew and {Oh}, S. Peng},
        title = "{Lyman {\ensuremath{\alpha}} radiative transfer in a multiphase medium}",
      journal = {\mnras},
         year = "2006",
        month = "Apr",
       volume = {367},
       number = {3},
        pages = {979-1002},
          doi = {10.1111/j.1365-2966.2005.09870.x},
archivePrefix = {arXiv},
       eprint = {astro-ph/0507586},
 primaryClass = {astro-ph},
       adsurl = {https://ui.adsabs.harvard.edu/abs/2006MNRAS.367..979H}
}

@article{Unno1952,
author = {Unno, W.},
issn = {0004-6264},
journal = {PASJ},
pages = {100},
title = {{Note on the Zanstra Redistribution in Planetary Nebulae}},
url = {http://adsabs.harvard.edu/abs/1952PASJ....4..100U http://adsabs.harvard.edu/cgi-bin/nph-data{\_}query?bibcode=1952PASJ....4..100U{\&}link{\_}type=ARTICLE},
volume = {4},
year = {1952}
}

@article{Hummer1962,
author = {Hummer, D. G.},
doi = {10.1093/mnras/125.1.21},
issn = {0035-8711},
journal = {MNRAS},
pages = {21--37},
shorttitle = {Non-coherent scattering},
title = {{Non-coherent scattering: I. The redistribution function with Doppler broadening}},
url = {http://adsabs.harvard.edu/abs/1962MNRAS.125...21H http://adsabs.harvard.edu/cgi-bin/nph-data{\_}query?bibcode=1962MNRAS.125...21H{\&}link{\_}type=ARTICLE},
volume = {125},
year = {1962}
}

@article{Rybicki1994,
author = {Rybicki, George B. and Dell'Antonio, Ian P.},
doi = {10.1086/174170},
issn = {0004-637X},
journal = {ApJ},
month = {jun},
pages = {603--617},
title = {{The time development of a resonance line in the expanding universe}},
url = {http://adsabs.harvard.edu/abs/1994ApJ...427..603R},
volume = {427},
year = {1994}
}

@article{Tasitsiomi2006,
author = {Tasitsiomi, Argyro},
doi = {10.1086/504460},
issn = {0004-637X},
journal = {ApJ},
month = jul,
pages = {792--813},
title = {{Ly$\alpha$ Radiative Transfer in Cosmological Simulations and Application to a z \~{}= 8 Ly$\alpha$ Emitter}},
url = {http://adsabs.harvard.edu/abs/2006ApJ...645..792T http://adsabs.harvard.edu/cgi-bin/nph-data\_query?bibcode=2006ApJ...645..792T\&link\_type=ARTICLE},
volume = {645},
year = {2006}
}

@ARTICLE{Tasitsiomi2006b,
       author = {{Tasitsiomi}, Argyro},
        title = "{On the Transfer of Resonant-Line Radiation in Mesh Simulations}",
      journal = {\apj},
         year = 2006,
        month = sep,
       volume = {648},
       number = {1},
        pages = {762-766},
          doi = {10.1086/505682},
archivePrefix = {arXiv},
       eprint = {astro-ph/0601562},
 primaryClass = {astro-ph},
       adsurl = {https://ui.adsabs.harvard.edu/abs/2006ApJ...648..762T}
}

@ARTICLE{Higgins2012,
       author = {{Higgins}, Jonathan and {Meiksin}, Avery},
        title = "{The scattering of Ly{\ensuremath{\alpha}} radiation in the intergalactic medium: numerical methods and solutions}",
      journal = {\mnras},
         year = 2012,
        month = nov,
       volume = {426},
       number = {3},
        pages = {2380-2403},
          doi = {10.1111/j.1365-2966.2012.21917.x},
archivePrefix = {arXiv},
       eprint = {1206.6759},
 primaryClass = {astro-ph.CO},
       adsurl = {https://ui.adsabs.harvard.edu/abs/2012MNRAS.426.2380H}
}

@article{Laursen2009,
author = {Laursen, Peter and Razoumov, Alexei O. and Sommer-Larsen, Jesper},
doi = {10.1088/0004-637X/696/1/853},
issn = {0004-637X},
journal = {ApJ},
month = may,
pages = {853--869},
title = {{Ly$\alpha$ Radiative Transfer in Cosmological Simulations Using Adaptive Mesh Refinement}},
url = {http://adsabs.harvard.edu/abs/2009ApJ...696..853L http://adsabs.harvard.edu/cgi-bin/nph-data\_query?bibcode=2009ApJ...696..853L\&link\_type=ARTICLE},
volume = {696},
year = {2009}
}

@ARTICLE{Smith2018,
       author = {{Smith}, Aaron and {Tsang}, Benny T. -H. and {Bromm}, Volker and
         {Milosavljevi{\'c}}, Milo{\v{s}}},
        title = "{Discrete diffusion Lyman {\ensuremath{\alpha}} radiative transfer}",
      journal = {\mnras},
         year = "2018",
        month = "Sep",
       volume = {479},
       number = {2},
        pages = {2065-2078},
          doi = {10.1093/mnras/sty1509},
archivePrefix = {arXiv},
       eprint = {1709.10187},
 primaryClass = {astro-ph.CO},
       adsurl = {https://ui.adsabs.harvard.edu/abs/2018MNRAS.479.2065S}
}

@ARTICLE{Smith2015,
       author = {{Smith}, Aaron and {Safranek-Shrader}, Chalence and {Bromm}, Volker and
         {Milosavljevi{\'c}}, Milo{\v{s}}},
        title = "{The Lyman {\ensuremath{\alpha}} signature of the first galaxies}",
      journal = {\mnras},
         year = "2015",
        month = "Jun",
       volume = {449},
       number = {4},
        pages = {4336-4362},
          doi = {10.1093/mnras/stv565},
archivePrefix = {arXiv},
       eprint = {1409.4480},
 primaryClass = {astro-ph.CO},
       adsurl = {https://ui.adsabs.harvard.edu/abs/2015MNRAS.449.4336S}
}

@ARTICLE{Smith2019,
       author = {{Smith}, Aaron and {Ma}, Xiangcheng and {Bromm}, Volker and
         {Finkelstein}, Steven L. and {Hopkins}, Philip F. and
         {Faucher-Gigu{\`e}re}, Claude-Andr{\'e} and {Kere{\v{s}}}, Du{\v{s}}an},
        title = "{The physics of Lyman {\ensuremath{\alpha}} escape from high-redshift galaxies}",
      journal = {\mnras},
         year = 2019,
        month = mar,
       volume = {484},
       number = {1},
        pages = {39-59},
          doi = {10.1093/mnras/sty3483},
archivePrefix = {arXiv},
       eprint = {1810.08185},
 primaryClass = {astro-ph.GA},
       adsurl = {https://ui.adsabs.harvard.edu/abs/2019MNRAS.484...39S}
}

@ARTICLE{Michel-Dansac2020,
       author = {{Michel-Dansac}, L. and {Blaizot}, J. and {Garel}, T. and
         {Verhamme}, A. and {Kimm}, T. and {Trebitsch}, M.},
        title = "{RASCAS: RAdiation SCattering in Astrophysical Simulations}",
      journal = {\aap},
         year = 2020,
        month = mar,
       volume = {635},
          eid = {A154},
        pages = {A154},
          doi = {10.1051/0004-6361/201834961},
archivePrefix = {arXiv},
       eprint = {2001.11252},
 primaryClass = {astro-ph.GA},
       adsurl = {https://ui.adsabs.harvard.edu/abs/2020A&A...635A.154M}
}

@ARTICLE{GeWise2017,
       author = {{Ge}, Qi and {Wise}, John H.},
        title = "{On the effect of Lyman {\ensuremath{\alpha}} trapping during the initial collapse of massive black hole seeds}",
      journal = {\mnras},
         year = 2017,
        month = dec,
       volume = {472},
       number = {3},
        pages = {2773-2786},
          doi = {10.1093/mnras/stx2074},
archivePrefix = {arXiv},
       eprint = {1706.01467},
 primaryClass = {astro-ph.GA},
       adsurl = {https://ui.adsabs.harvard.edu/abs/2017MNRAS.472.2773G}
}

@ARTICLE{Smith2017,
       author = {{Smith}, Aaron and {Bromm}, Volker and {Loeb}, Abraham},
        title = "{Lyman {\ensuremath{\alpha}} radiation hydrodynamics of galactic winds before cosmic reionization}",
      journal = {\mnras},
         year = 2017,
        month = jan,
       volume = {464},
       number = {3},
        pages = {2963-2978},
          doi = {10.1093/mnras/stw2591},
archivePrefix = {arXiv},
       eprint = {1607.07166},
 primaryClass = {astro-ph.GA},
       adsurl = {https://ui.adsabs.harvard.edu/abs/2017MNRAS.464.2963S}
}

@ARTICLE{Seon2020,
       author = {{Seon}, Kwang-il and {Kim}, Chang-Goo},
        title = "{Ly{\ensuremath{\alpha}} Radiative Transfer: Monte Carlo Simulation of the Wouthuysen-Field Effect}",
      journal = {\apjs},
         year = 2020,
        month = sep,
       volume = {250},
       number = {1},
          eid = {9},
        pages = {9},
          doi = {10.3847/1538-4365/aba2d6},
archivePrefix = {arXiv},
       eprint = {2005.00238},
 primaryClass = {astro-ph.GA},
       adsurl = {https://ui.adsabs.harvard.edu/abs/2020ApJS..250....9S}
}

@ARTICLE{Tomaselli2021,
       author = {{Tomaselli}, G.~M. and {Ferrara}, A.},
        title = "{Lyman-alpha radiation pressure: an analytical exploration}",
      journal = {\mnras},
         year = 2021,
        month = jun,
       volume = {504},
       number = {1},
        pages = {89-100},
          doi = {10.1093/mnras/stab876},
archivePrefix = {arXiv},
       eprint = {2103.14655},
 primaryClass = {astro-ph.GA},
       adsurl = {https://ui.adsabs.harvard.edu/abs/2021MNRAS.504...89T}
}

@ARTICLE{LaoSmith2020,
       author = {{Lao}, Bing-Xin and {Smith}, Aaron},
        title = "{Resonant-line radiative transfer within power-law density profiles}",
      journal = {\mnras},
         year = 2020,
        month = sep,
       volume = {497},
       number = {3},
        pages = {3925-3942},
          doi = {10.1093/mnras/staa2198},
archivePrefix = {arXiv},
       eprint = {2005.09692},
 primaryClass = {astro-ph.GA},
       adsurl = {https://ui.adsabs.harvard.edu/abs/2020MNRAS.497.3925L}
}

@ARTICLE{Basko1981,
       author = {{Basko}, M.~M.},
        title = "{Thermalization Length of Resonance Radiation in the Case of Partial Frequency Redistribution}",
      journal = {Astrophysics},
         year = 1981,
        month = jan,
       volume = {17},
        pages = {69},
          doi = {10.1007/BF01014298},
       adsurl = {https://ui.adsabs.harvard.edu/abs/1981Ap.....17...69B}
}

@ARTICLE{Rybicki2006,
       author = {{Rybicki}, George B.},
        title = "{Improved Fokker-Planck Equation for Resonance-Line Scattering}",
      journal = {\apj},
         year = 2006,
        month = aug,
       volume = {647},
       number = {1},
        pages = {709-718},
          doi = {10.1086/505327},
archivePrefix = {arXiv},
       eprint = {astro-ph/0603047},
 primaryClass = {astro-ph},
       adsurl = {https://ui.adsabs.harvard.edu/abs/2006ApJ...647..709R}
}

@ARTICLE{Verhamme2006,
       author = {{Verhamme}, A. and {Schaerer}, D. and {Maselli}, A.},
        title = "{3D Ly{\ensuremath{\alpha}} radiation transfer. I. Understanding Ly{\ensuremath{\alpha}} line profile morphologies}",
      journal = {\aap},
         year = 2006,
        month = dec,
       volume = {460},
       number = {2},
        pages = {397-413},
          doi = {10.1051/0004-6361:20065554},
archivePrefix = {arXiv},
       eprint = {astro-ph/0608075},
 primaryClass = {astro-ph},
       adsurl = {https://ui.adsabs.harvard.edu/abs/2006A&A...460..397V}
}

@ARTICLE{Semelin2007,
       author = {{Semelin}, B. and {Combes}, F. and {Baek}, S.},
        title = "{Lyman-alpha radiative transfer during the epoch of reionization: contribution to 21-cm signal fluctuations}",
      journal = {\aap},
         year = 2007,
        month = nov,
       volume = {474},
       number = {2},
        pages = {365-374},
          doi = {10.1051/0004-6361:20077965},
archivePrefix = {arXiv},
       eprint = {0707.2483},
 primaryClass = {astro-ph},
       adsurl = {https://ui.adsabs.harvard.edu/abs/2007A&A...474..365S}
}

@ARTICLE{Yajima2012,
       author = {{Yajima}, Hidenobu and {Li}, Yuexing and {Zhu}, Qirong and {Abel}, Tom},
        title = "{ART$^{2}$: coupling Ly{\ensuremath{\alpha}} line and multi-wavelength continuum radiative transfer}",
      journal = {\mnras},
         year = 2012,
        month = aug,
       volume = {424},
       number = {2},
        pages = {884-901},
          doi = {10.1111/j.1365-2966.2012.21228.x},
archivePrefix = {arXiv},
       eprint = {1109.4891},
 primaryClass = {astro-ph.CO},
       adsurl = {https://ui.adsabs.harvard.edu/abs/2012MNRAS.424..884Y}
}

@ARTICLE{GronkeDijkstra2016,
       author = {{Gronke}, M. and {Dijkstra}, M.},
        title = "{Lyman-{\ensuremath{\alpha}} Spectra from Multiphase Outflows, and their Connection to Shell Models}",
      journal = {\apj},
         year = 2016,
        month = jul,
       volume = {826},
       number = {1},
          eid = {14},
        pages = {14},
          doi = {10.3847/0004-637X/826/1/14},
archivePrefix = {arXiv},
       eprint = {1604.06805},
 primaryClass = {astro-ph.GA},
       adsurl = {https://ui.adsabs.harvard.edu/abs/2016ApJ...826...14G}
}

@ARTICLE{SmithMW2022,
       author = {{Smith}, Aaron and {Kannan}, Rahul and {Tacchella}, Sandro and {Vogelsberger}, Mark and {Hernquist}, Lars and {Marinacci}, Federico and {Sales}, Laura V. and {Torrey}, Paul and {Li}, Hui and {Yeh}, Jessica Y. -C. and {Qi}, Jia},
        title = "{The physics of Lyman-{\ensuremath{\alpha}} escape from disc-like galaxies}",
      journal = {\mnras},
         year = 2022,
        month = nov,
       volume = {517},
       number = {1},
        pages = {1-27},
          doi = {10.1093/mnras/stac2641},
archivePrefix = {arXiv},
       eprint = {2111.13721},
 primaryClass = {astro-ph.GA},
       adsurl = {https://ui.adsabs.harvard.edu/abs/2022MNRAS.517....1S}
}

@ARTICLE{Nebrin2025,
       author = {{Nebrin}, Olof and {Smith}, Aaron and {Lorinc}, Kevin and {H{\"o}rnquist}, Johan and {Larson}, {\r{A}}sa and {Mellema}, Garrelt and {Giri}, Sambit K.},
        title = "{Lyman-{\ensuremath{\alpha}} feedback prevails at Cosmic Dawn: implications for the first galaxies, stars, and star clusters}",
      journal = {\mnras},
         year = 2025,
        month = feb,
       volume = {537},
       number = {2},
        pages = {1646-1687},
          doi = {10.1093/mnras/staf038},
archivePrefix = {arXiv},
       eprint = {2409.19288},
 primaryClass = {astro-ph.GA},
       adsurl = {https://ui.adsabs.harvard.edu/abs/2025MNRAS.537.1646N}
}

@ARTICLE{McClellan2022,
       author = {{McClellan}, B. Connor and {Davis}, Shane W. and {Arras}, Phil},
        title = "{A Novel Solution for Resonant Scattering Using Self-consistent Boundary Conditions}",
      journal = {\apj},
         year = 2022,
        month = jul,
       volume = {934},
       number = {1},
          eid = {37},
        pages = {37},
          doi = {10.3847/1538-4357/ac7724},
archivePrefix = {arXiv},
       eprint = {2205.05082},
 primaryClass = {astro-ph.EP},
       adsurl = {https://ui.adsabs.harvard.edu/abs/2022ApJ...934...37M}
}

@ARTICLE{Kimm2018,
       author = {{Kimm}, Taysun and {Haehnelt}, Martin and {Blaizot}, J{\'e}r{\'e}my and {Katz}, Harley and {Michel-Dansac}, L{\'e}o and {Garel}, Thibault and {Rosdahl}, Joakim and {Teyssier}, Romain},
        title = "{Impact of Lyman alpha pressure on metal-poor dwarf galaxies}",
      journal = {\mnras},
         year = 2018,
        month = apr,
       volume = {475},
       number = {4},
        pages = {4617-4635},
          doi = {10.1093/mnras/sty126},
archivePrefix = {arXiv},
       eprint = {1801.04952},
 primaryClass = {astro-ph.GA},
       adsurl = {https://ui.adsabs.harvard.edu/abs/2018MNRAS.475.4617K}
}

@ARTICLE{Smith2025,
       author = {{Smith}, Aaron and {Lorinc}, Kevin and {Nebrin}, Olof and {Lao}, Bing-Xin},
        title = "{Lyman-{\ensuremath{\alpha}} resonant-line radiative transfer in expanding media}",
      journal = {\mnras},
         year = 2025,
        month = jun,
          doi = {10.1093/mnras/staf961},
archivePrefix = {arXiv},
       eprint = {2501.01928},
primaryClass = {astro-ph.GA},
       adsurl = {https://ui.adsabs.harvard.edu/abs/2025MNRAS.tmp..933S}
}

@ARTICLE{Smith2020,
       author = {{Smith}, Aaron and {Kannan}, Rahul and {Tsang}, Benny T. -H. and {Vogelsberger}, Mark and {Pakmor}, R{\"u}diger},
        title = "{AREPO-MCRT: Monte Carlo Radiation Hydrodynamics on a Moving Mesh}",
      journal = {\apj},
         year = 2020,
        month = dec,
       volume = {905},
       number = {1},
          eid = {27},
        pages = {27},
          doi = {10.3847/1538-4357/abc47e},
archivePrefix = {arXiv},
       eprint = {2008.01750},
 primaryClass = {astro-ph.GA},
       adsurl = {https://ui.adsabs.harvard.edu/abs/2020ApJ...905...27S}
}

@ARTICLE{Camps2018,
       author = {{Camps}, Peter and {Baes}, Maarten},
        title = "{The Failure of Monte Carlo Radiative Transfer at Medium to High Optical Depths}",
      journal = {\apj},
         year = 2018,
        month = jul,
       volume = {861},
       number = {2},
          eid = {80},
        pages = {80},
          doi = {10.3847/1538-4357/aac824},
archivePrefix = {arXiv},
       eprint = {1805.09502},
 primaryClass = {astro-ph.IM},
       adsurl = {https://ui.adsabs.harvard.edu/abs/2018ApJ...861...80C}
}

@ARTICLE{Meiksin2006,
       author = {{Meiksin}, Avery},
        title = "{Energy transfer by the scattering of resonant photons}",
      journal = {\mnras},
         year = 2006,
        month = aug,
       volume = {370},
       number = {4},
        pages = {2025-2037},
          doi = {10.1111/j.1365-2966.2006.10632.x},
archivePrefix = {arXiv},
       eprint = {astro-ph/0603855},
 primaryClass = {astro-ph},
       adsurl = {https://ui.adsabs.harvard.edu/abs/2006MNRAS.370.2025M}
}

@ARTICLE{DijkstraLoeb2008,
       author = {{Dijkstra}, Mark and {Loeb}, Abraham},
        title = "{Ly{\ensuremath{\alpha}}-driven outflows around star-forming galaxies}",
      journal = {\mnras},
         year = 2008,
        month = nov,
       volume = {391},
       number = {1},
        pages = {457-466},
          doi = {10.1111/j.1365-2966.2008.13920.x},
archivePrefix = {arXiv},
       eprint = {0807.2645},
 primaryClass = {astro-ph},
       adsurl = {https://ui.adsabs.harvard.edu/abs/2008MNRAS.391..457D}
}

@ARTICLE{Abe2018,
       author = {{Abe}, Makito and {Yajima}, Hidenobu},
        title = "{Suppression of globular cluster formation in metal-poor gas clouds by Lyman {\ensuremath{\alpha}} radiation feedback}",
      journal = {\mnras},
         year = 2018,
        month = mar,
       volume = {475},
       number = {1},
        pages = {L130-L134},
          doi = {10.1093/mnrasl/sly018},
archivePrefix = {arXiv},
       eprint = {1801.10473},
 primaryClass = {astro-ph.GA},
       adsurl = {https://ui.adsabs.harvard.edu/abs/2018MNRAS.475L.130A}
}

@ARTICLE{Heinzel1978,
       author = {{Heinzel}, P.},
        title = "{Derivatives of the Voigt Functions}",
      journal = {Bulletin of the Astronomical Institutes of Czechoslovakia},
         year = 1978,
        month = jan,
       volume = {29},
        pages = {159},
       adsurl = {https://ui.adsabs.harvard.edu/abs/1978BAICz..29..159H}
}

@ARTICLE{Risken1991,
       author = {{Risken}, H. and {Caugheyz}, T.~K.},
        title = "{The Fokker-Planck Equation: Methods of Solution and Application, 2nd ed.}",
      journal = {Journal of Applied Mechanics},
         year = 1991,
        month = jan,
       volume = {58},
       number = {3},
        pages = {860},
          doi = {10.1115/1.2897281},
       adsurl = {https://ui.adsabs.harvard.edu/abs/1991JAM....58..860R}
}

@ARTICLE{Lucy1999,
       author = {{Lucy}, L.~B.},
        title = "{Computing radiative equilibria with Monte Carlo techniques}",
      journal = {\aap},
         year = 1999,
        month = apr,
       volume = {344},
        pages = {282-288},
       adsurl = {https://ui.adsabs.harvard.edu/abs/1999A&A...344..282L}
}

@BOOK{Hubeny2015,
       author = {{Hubeny}, Ivan and {Mihalas}, Dimitri},
        title = "{Theory of Stellar Atmospheres. An Introduction to Astrophysical Non-equilibrium Quantitative Spectroscopic Analysis}",
         year = 2015,
       adsurl = {https://ui.adsabs.harvard.edu/abs/2015tsaa.book.....H},
      publisher = {Princeton University Press}
}

@ARTICLE{Ayres1985,
       author = {{Ayres}, T.~R.},
        title = "{A physically realistic approximate form for the redistribution function R(II-A)}",
      journal = {\apj},
         year = 1985,
        month = jul,
       volume = {294},
        pages = {153-157},
          doi = {10.1086/163283},
       adsurl = {https://ui.adsabs.harvard.edu/abs/1985ApJ...294..153A}
}

@ARTICLE{SmithDCBH2017,
       author = {{Smith}, Aaron and {Becerra}, Fernando and {Bromm}, Volker and {Hernquist}, Lars},
        title = "{Radiative effects during the assembly of direct collapse black holes}",
      journal = {\mnras},
         year = 2017,
        month = nov,
       volume = {472},
       number = {1},
        pages = {205-216},
          doi = {10.1093/mnras/stx1993},
archivePrefix = {arXiv},
       eprint = {1706.02751},
 primaryClass = {astro-ph.GA},
       adsurl = {https://ui.adsabs.harvard.edu/abs/2017MNRAS.472..205S}
}

@ARTICLE{Smith2016,
       author = {{Smith}, Aaron and {Bromm}, Volker and {Loeb}, Abraham},
        title = "{Evidence for a direct collapse black hole in the Lyman {\ensuremath{\alpha}} source CR7}",
      journal = {\mnras},
         year = 2016,
        month = aug,
       volume = {460},
       number = {3},
        pages = {3143-3151},
          doi = {10.1093/mnras/stw1129},
archivePrefix = {arXiv},
       eprint = {1602.07639},
 primaryClass = {astro-ph.GA},
       adsurl = {https://ui.adsabs.harvard.edu/abs/2016MNRAS.460.3143S}
}

@ARTICLE{Mushano2024,
       author = {{Mushano}, Takuya and {Ogawa}, Takumi and {Ohsuga}, Ken and {Yajima}, Hidenobu and {Omukai}, Kazuyuki},
        title = "{Impact of the Ly{\ensuremath{\alpha}} radiation force on super-Eddington accretion on to a massive black hole}",
      journal = {\pasj},
         year = 2024,
        month = dec,
       volume = {76},
       number = {6},
        pages = {1260-1269},
          doi = {10.1093/pasj/psae086},
archivePrefix = {arXiv},
       eprint = {2410.04378},
 primaryClass = {astro-ph.HE},
       adsurl = {https://ui.adsabs.harvard.edu/abs/2024PASJ...76.1260M}
}

@ARTICLE{Byrohl2025,
       author = {{Byrohl}, Chris and {Nelson}, Dylan},
        title = "{THOR: a GPU-accelerated and MPI-parallel radiative transfer code}",
      journal = {arXiv e-prints},
         year = 2025,
        month = jul,
          eid = {arXiv:2507.11603},
        pages = {arXiv:2507.11603},
          doi = {10.48550/arXiv.2507.11603},
archivePrefix = {arXiv},
       eprint = {2507.11603},
 primaryClass = {astro-ph.GA},
       adsurl = {https://ui.adsabs.harvard.edu/abs/2025arXiv250711603B}
}

@ARTICLE{LiangOh2025,
       author = {{Liang}, Naixin and {Oh}, Siang Peng},
        title = "{L{\'e}vy Flights and Leaky Boxes: Anomalous Diffusion of Cosmic Rays}",
      journal = {arXiv e-prints},
         year = 2025,
        month = mar,
          eid = {arXiv:2503.10747},
        pages = {arXiv:2503.10747},
          doi = {10.48550/arXiv.2503.10747},
archivePrefix = {arXiv},
       eprint = {2503.10747},
 primaryClass = {astro-ph.HE},
       adsurl = {https://ui.adsabs.harvard.edu/abs/2025arXiv250310747L}
}

@ARTICLE{Metzler2000,
       author = {{Metzler}, Ralf and {Klafter}, Joseph},
        title = "{The random walk's guide to anomalous diffusion: a fractional dynamics approach}",
      journal = {\physrep},
         year = 2000,
        month = dec,
       volume = {339},
       number = {1},
        pages = {1-77},
          doi = {10.1016/S0370-1573(00)00070-3},
       adsurl = {https://ui.adsabs.harvard.edu/abs/2000PhR...339....1M}
}

@ARTICLE{Ferrara2025,
       author = {{Ferrara}, Andrea and {Manzoni}, Daniele and {Ntormousi}, Evangelia},
        title = "{Is feedback-free star formation possible?}",
      journal = {The Open Journal of Astrophysics},
         year = 2025,
        month = sep,
       volume = {8},
          eid = {140},
        pages = {140},
          doi = {10.33232/001c.144792},
archivePrefix = {arXiv},
       eprint = {2509.02566},
 primaryClass = {astro-ph.GA},
       adsurl = {https://ui.adsabs.harvard.edu/abs/2025OJAp....8E.140F}
}

@ARTICLE{Manzoni2025,
       author = {{Manzoni}, D. and {Ferrara}, A.},
        title = "{Lyman-$α$ radiation pressure regulates star formation efficiency}",
      journal = {arXiv e-prints},
         year = 2025,
        month = oct,
          eid = {arXiv:2510.25950},
        pages = {arXiv:2510.25950},
          doi = {10.48550/arXiv.2510.25950},
archivePrefix = {arXiv},
       eprint = {2510.25950},
 primaryClass = {astro-ph.GA},
       adsurl = {https://ui.adsabs.harvard.edu/abs/2025arXiv251025950M}
}

@ARTICLE{McClymont2025,
       author = {{McClymont}, William and {Smith}, Aaron and {Tacchella}, Sandro},
        title = "{Modelling the nebular emission of galaxies across cosmic time with COLT}",
      journal = {arXiv e-prints},
         year = 2025,
        month = oct,
          eid = {arXiv:2510.13952},
        pages = {arXiv:2510.13952},
          doi = {10.48550/arXiv.2510.13952},
archivePrefix = {arXiv},
       eprint = {2510.13952},
 primaryClass = {astro-ph.GA},
       adsurl = {https://ui.adsabs.harvard.edu/abs/2025arXiv251013952M}
}

@ARTICLE{AlmadaMonter2025,
       author = {{Almada Monter}, Silvia and {Gronke}, Max and {Chang}, Seok-Jun},
        title = "{Lyman-$α$ Escape through Anisotropic Media}",
      journal = {arXiv e-prints},
         year = 2025,
        month = sep,
          eid = {arXiv:2509.19184},
        pages = {arXiv:2509.19184},
          doi = {10.48550/arXiv.2509.19184},
archivePrefix = {arXiv},
       eprint = {2509.19184},
 primaryClass = {astro-ph.GA},
       adsurl = {https://ui.adsabs.harvard.edu/abs/2025arXiv250919184A}
}

@ARTICLE{Roth2015,
       author = {{Roth}, Nathaniel and {Kasen}, Daniel},
        title = "{Monte Carlo Radiation-Hydrodynamics With Implicit Methods}",
      journal = {\apjs},
         year = 2015,
        month = mar,
       volume = {217},
       number = {1},
          eid = {9},
        pages = {9},
          doi = {10.1088/0067-0049/217/1/9},
archivePrefix = {arXiv},
       eprint = {1404.4652},
 primaryClass = {astro-ph.IM},
       adsurl = {https://ui.adsabs.harvard.edu/abs/2015ApJS..217....9R}
}

%%%%%%%%%%%%%%%%%%%%%%%%%%%%%%%%%%%%%%%%%%%%%%%%%%

%%%%%%%%%%%%%%%%% APPENDICES %%%%%%%%%%%%%%%%%%%%%

% If you want to present additional material which would interrupt the flow of the main paper, it can be placed in an Appendix which appears after the list of references.

\appendix

\section{Redistribution function moments} \label{sec:rf_moments}
We consider the moments of the redistribution term $k_{x'} R_{x' \rightarrow x}$ needed in the derivation of the Fokker--Planck approximation. For convenience, we use the normalized Voigt profile $\varphi(x) = H(a,x)/\sqrt\pi$ where we drop the $a$ dependence since it isl constant under our isothermal regime. The moments of the redistribution function are defined as
\begin{equation}
    \mathcal{M}_k(\bm{n'}, x, \bm{n}) \equiv \int (x' - x)^k R_{\text{II}}(x', \bm{n}';x,\bm{n}) \,\dd x' \, .
\end{equation}
For the cases we consider, the angular dependence reduces to a dependence on $\mu \equiv \bm{n} \cdot \bm{n}'$. The easiest way to evaluate these integrals is in Fourier space based on the Fourier transform of $R_\text{II}$ \citep{Rybicki1994}. Here we ignore the recoil term and the additional effects considered in \citet{Basko1981}, \citet{Rybicki2006}, and  \citet{Meiksin2006}. In addition, we only require the angular-averaged moments, so we define $\mathcal{M}_k(x) \equiv \iint\mathcal{M}_k(\bm{n'}, x, \bm{n}) \dd\Omega'\dd\Omega$. To find these angular-averaged moments, we integrate the results with angular dependence. The atom-frame phase function is either $g_A(\bm{n}',\bm{n}) = 1$ in the isotropic case or $g_B(\bm{n'},\bm{n}) = \frac{3}{4}[1 + (\bm{n}'\cdot\bm{n})^2]$ in the dipole scattering case, or simply $g_B(\mu) = \frac{3}{4}(1 + \mu^2)$. The first five fully angle-dependent moments are \citep[for up to second order, see][]{Rybicki1994}:
\begin{align}
    \mathcal{M}_0(x,\mu) &= g(\mu)\varphi(x) \\
    \mathcal{M}_1(x,\mu) &= \frac{1}{2} g(\mu) (1 - \mu) \varphi'(x) \\
    \mathcal{M}_2(x,\mu) &= g(\mu) \left[(1 - \mu)\varphi(x) + \frac{1}{4} (1 - \mu)^2 \varphi''(x)\right] \label{eq:rii_M2}\\
    \mathcal{M}_3(x,\mu) &= g(\mu) \left[-\frac{3}{2} ( 1 - \mu)^2 \varphi'(x) + \frac{1}{8}(1 - \mu)^3 \varphi'''(x)\right]\\
    \mathcal{M}_4(x,\mu) &= g(\mu)\bigg[3(1 - \mu)^2 \varphi(x) \notag\\&\qquad- \frac{3}{2} ( 1 - \mu)^3 \varphi''(x) + \frac{1}{16}(1 - \mu)^4 \varphi''''(x)\bigg]\, .
\end{align}
To proceed, we need the moments of the phase functions, which we define as $G_k = \frac{1}{2}\int_{-1}^1 g(\mu)\,\dd\mu$, and they are
\begin{equation}
    G_{\{0,2,4\}}^A = \left\{1, \frac{1}{3}, \frac{1}{5}\right\} \quad \text{and} \quad G_{\{0,2,4\}}^B = \left\{1, \frac{2}{5}, \frac{9}{35}\right\} \, ,
\end{equation}
where the odd moments are $0$, and the superscript denotes whether its case A (isotropic) or case B (dipole) scattering. We leave it general (no superscript) below so both can be easily incorporated. We find (see \citet{Meiksin2006} for the isotropic scattering case with recoil terms, noting differences in convention)
\begin{align}
    \mathcal{M}_0(x) &= \varphi(x) \\
    \mathcal{M}_1(x) &= \frac{1}{2} \varphi'(x) \\
    \mathcal{M}_2(x) &= \varphi(x) + \frac{1}{4} ( 1 + G_2) \varphi''(x) \\
    \mathcal{M}_3(x) &= -\frac{3}{2} (1 + G_2) \varphi'(x) + \frac{1}{8}(1 +3 G_2) \varphi'''(x)\\
    \mathcal{M}_4(x) &= 3 ( 1 + G_2) \varphi(x) - \frac{3}{2} (1 + 3 G_2)\varphi''(x)\notag \\&\qquad+ \frac{1}{16} ( 1 + 6 G_2 + G_4) \varphi''''(x) \,.
\end{align}
The derivatives of the Voigt profile are more easily expressed in terms of the Voigt functions $H(a,x)$, which is defined in Eq.~(\ref{eq:H}), and $K(a,x)$, which are defined as
\begin{equation}
    K(a,x) = \frac{1}{\pi} \int_{-\infty}^\infty \frac{e^{-y^2} (x -y)}{(y-x)^2 + a^2} \,\dd y\, .
\end{equation}
\begin{figure}
    \centering
    \includegraphics[width=\columnwidth]{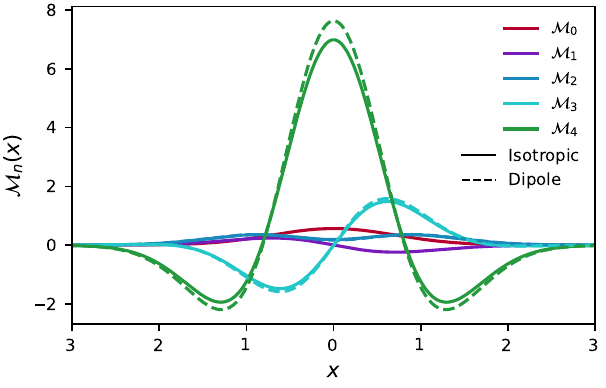}
    \caption{The first five angular-averaged moments of the $R_\text{II}$ redistribution function $\mathcal{M}_k$. The solid lines show the moments assuming the isotropic scattering case and the dotted lines show the dipole scattering case.}
    \label{fig:rii_moments}
\end{figure}
We can then use the recurrence relations derived in \citet{Heinzel1978} to write the above moments exactly in terms of the Voigt functions. The only derivative that does not appear explicitly in \citet{Heinzel1978} is the fourth derivative term which we can write as 
\begin{align}
    &\frac{\partial^4 H}{\partial x^4} = \,4H(a,x) \left[3 + 4a^4 - 12 x^2 + 4x^4 + 12 a^2 ( 1 - 2x^2)\right] \notag\\
    &\quad+ 32 a x K(a,x) \left(2a^2 - 2x^2 + 3\right) + \frac{8 a}{\sqrt\pi} \left(6x^2 -2 a^2 - 5\right) \, .
\end{align}
In Fig.~\ref{fig:rii_moments}, we plot the first five moments of the $R_\text{II}$ redistribution function using the joint-probability definition.  There is a considerable difference for higher moments between the isotropic and dipole scattering cases, while the first few moments are identical (or nearly identical in the $n = 2$ case. 
\begin{figure}
    \centering
    \includegraphics[width=\columnwidth]{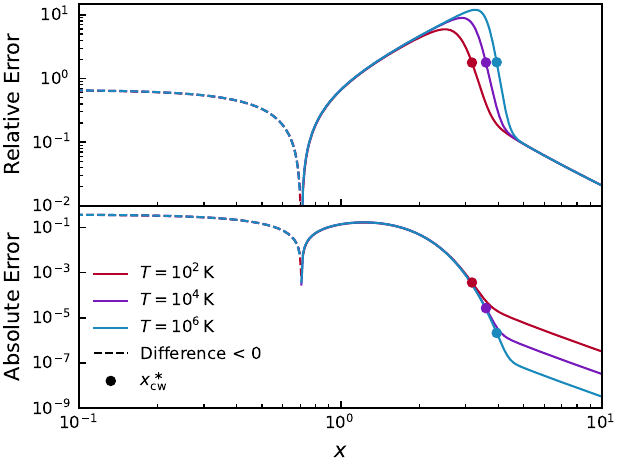}
    \caption{The Error in the redistribution moment approximation for various temperatures. We plot only the isotropic case since the dipole case is nearly identical for $\mathcal{M}_2$. The dots represent the value of $x_\text{cw}^\ast$ corresponding to the temperature plotted, which varies slightly over this temperature range. The agreement is only good in the wing. \textbf{Bottom Panel}: The absolute difference between the second moment and zeroth moments $|\mathcal{M}_2 - \mathcal{M}_0|$. \textbf{Top Panel}: the relative difference between the first and second moments $|\mathcal{M}_2 - \mathcal{M}_0|/ \mathcal{M}_0$.}
    \label{fig:rii_error}
\end{figure}
In \citet{Rybicki1994}, it was argued that dropping the second-order Voigt profile derivative term in Eq.~(\ref{eq:rii_M2}) was an acceptable approximation in the wing, e.g. $\mathcal{M}_2(x) \approx g(\mu)(1-\mu)\varphi(x)$, since it was the simplest choice to enforce photon conservation. However, the error in the core was simply accepted. After taking the angular average, the error in this approximation is equivalent to $|\mathcal{M}_2(x) - \mathcal{M}_0(x)|$, which we plot in Fig.~\ref{fig:rii_error} for a realistic range of temperatures. The error is low and falls off as a power law in the wing, but is quite high in the core (tens of percent), which explains the breakdown of the Fokker--Planck approximation for frequency-dependent quantities in the core. Making other approximations using only the first three moments will not lead to any additional accuracy, so the only way to improve the accuracy in the core with the Fokker--Planck method would be to use the higher-order moments in the Fokker--Planck expansion to account for the differences in frequency being on the order of the frequency itself. However, it is not obvious how to achieve this numerically while enforcing photon conservation.

\section{Return to the core} \label{sec:return_to_core}

\begin{figure}
    \centering
    \includegraphics[width=\columnwidth]{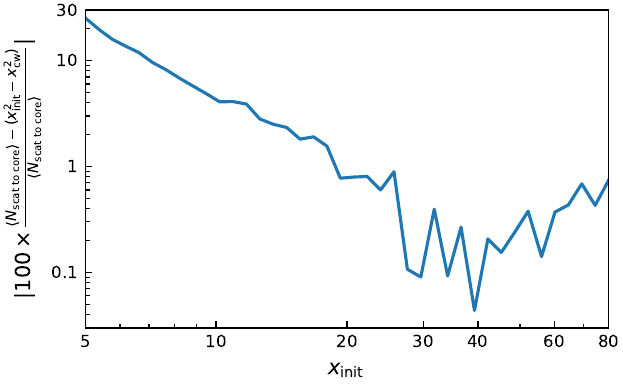}
    \caption{The relative error in the number of scatterings a photon takes to return to the core given an initial frequency $x_\text{init}$ when compared to the analytical estimate from Eq.~(\ref{eq:n_scat_to_core_analytical}). The error starts out high but decreases with a power-law scaling until the error $\leq 1 \%$. The MCRT simulations are with $5 \times 10^7$ photons for each initial frequency with 40 initial frequencies ranging from $x = 5$ to $x = 80$. The temperature is held fixed at $T = 10^4 \,\K$}
    \label{fig:n_scat_to_core_error}
\end{figure}

It is a well known scaling argument that for a photon in the wing ($x \gg x_\text{cw}^\ast$), it takes $x^2$ scatterings to return to the core. To see how this arises rigorously, we look at this process as the evolution of the frequency density distribution considering only the scattering process. Let $p_n(x)\,\dd x$ be the probability that a photon has a frequency between $x \in (x , x +\dd x)$ after n scattering events. Then the conditional redistribution function $R(x|x')$ (the probability that a photon is re-emitted at frequency x given that it was absorbed with frequency $x'$) allows us to write the recursive relationship
\begin{equation}
    p_{n+1}(x) = \int R(x|x')\,p_{n}(x')\,\dd x' \, .
\end{equation}
Subtracting $p_{n}(x)$ from both sides and letting $\Delta = x' - x$. There are enough scattering events that we can treat n as a continuous variable, so $p_{n+1}(x) - p_n(x) \rightarrow \frac{\partial p_n}{\partial n}$. Then expand $p_n(x+\Delta)$ and note that the resulting moments are equivalent to the moments in the appendix of \citet{Rybicki1994} normalized by the Voigt profile $\varphi(x)$. To second order,
\begin{equation}
    \frac{\partial p_n(x)}{\partial n} = M_1(x) \frac{\partial p_n(x)}{\partial x} + \frac{1}{2} M_2(x) \frac{\partial^2 p_n(x)}{\partial x^2} \, ,
\end{equation}
where
\begin{equation}
    M_k(x) = \int \Delta^k R(x | x+ \Delta) \,\dd \Delta \, .
\end{equation}
$M_1 = \varphi'(x)/\varphi(x) \approx - 1/x$ in the wing, and also since the second order term falls off as $\sim 6/x^2$, we can ignore it in the wing, so $M_2 \approx 1$. These moments are the drift and variance of $p_n$, respectively. This equation mirrors the backward Kolmogorov equation, whose Fokker--Planck operator can be used to solve for the mean hitting time \citep{Risken1991}. Let $T(x)$ be the mean number of scatterings to reach $x \in \{ |x| < x_{\text{cw}}^\ast\}$ given the initial frequency $x$. Then $T$ satisfies the equation
\begin{equation}
    \frac{1}{2} T''(x) -\frac{1}{x} T'(x) = -1
\end{equation}
subject to the boundary condition $T(x_{\text{cw}}^\ast) = 0$ and the natural condition $T''(x) < \infty$ as $x \rightarrow \infty$. This is physically motivated by escape to infinite frequency, suggested by an accelerating mean number of scatterings as a function of initial frequency, being non-physical in the context of resonant line scattering. The solution is
\begin{equation} \label{eq:n_scat_to_core_analytical}
    T(x) = x^2 - {x_\text{cw}^\ast}^2 \, ,
\end{equation}
which agrees with the back-of-the-envelope calculation in \citet{Osterbrock1962} with the additional boundary correction, previously absent in the Ly$\alpha$ literature.

\begin{figure}
    \centering
    \includegraphics[width=\columnwidth]{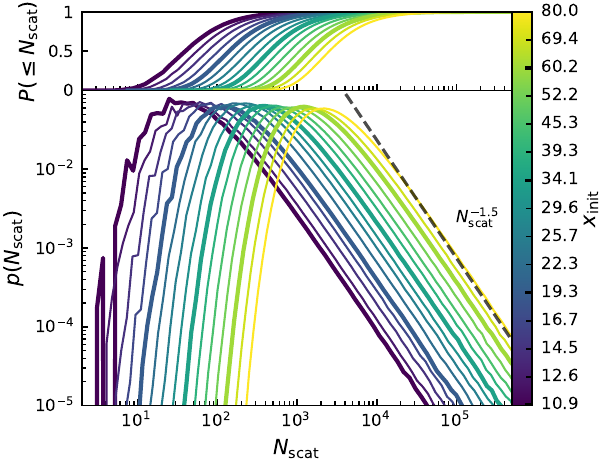}
    \caption{\textbf{Bottom:} The distribution $P(N_\text{scat})$ for various initial frequencies $x_\text{init}$ indicated by the color bar. The distributions shift over evenly indicating the power-law scaling of the mean. The distributions are fat-tailed on the upper end of $N_\text{scat}$ with a power-law slope of $\propto N_\text{scat}^{-1.5}$. \textbf{Top:} The cumulative distribution giving the probability that a photon undergoes $\leq N_\text{scat}$ scatterings to return to the core given $x_\text{init}$. The thicker lines correspond to thicker tick marks on the color bar and are meant to guide the eye. The simulations are the same as those described in Fig.~\ref{fig:n_scat_to_core_error}.}
    \label{fig:n_scat_to_core_distributions}
\end{figure}

To test this solution, we remove the spatial component of MCRT and simply scatter photons with a fixed initial frequency $x_\text{init}$ repeatedly until the frequency enters the core, so that $x < x_\text{cw}^\ast$. In Fig.~\ref{fig:n_scat_to_core_error}, we show the relative error in the analytical estimate from Eq.~(\ref{eq:n_scat_to_core_analytical}) and the MCRT simulations as a function of the initial frequency where all photons are injected. Once the initial frequency is far enough in the wing, the error becomes $\lesssim 1 \%$, and the correction for the boundary condition significantly reduces the error for wing frequencies closer to the core around $x \sim 10$. Fig.~\ref{fig:n_scat_to_core_distributions} shows the individual distributions for the number of scatterings that each photon took to reach the core from an initial frequency indicated by the color bar. The distributions simply shift, indicating the power-law scaling of the mean. The power-law fit indicates that this is a fat-tailed distribution, and the mean is significantly dragged up by photons that venture far into the wing ($x \geq 10^3$) before drifting back into the core (represented by the far right end of the tail). Numerically capturing the tail is difficult, and an escape condition was included for photons with $x \geq 10^4$ to avoid the photon drifting out towards infinity where the drift towards the core is 0. About $(5 \times 10^{-6}) \%$ across all initial frequencies actually escaped from this condition, but it explains the breakdown of the power-law scaling of the relative error for higher $x_\text{init}$ where the escape is more heavily weighted ($\sim (5 \times 10^{-5}) \%$ escape). The fit is empirical and, for now, we do not provide insight into where the scaling $-1.5$ comes from. The top panel shows the cumulative distribution, giving the probability that a photon takes $\leq N_\text{scat}$ scatterings to return to the core.

\end{document}